\documentclass[12pt]{article}
\usepackage{amsmath,amssymb,epsfig,graphicx}

\newcommand{\ket}[1]{|#1\rangle}

\begin{document}

\baselineskip=17pt

\begin{titlepage}
\rightline{\tt arXiv:1207.3335}
\rightline{\tt  UT-Komaba/12-6}
\begin{center}
\vskip 0.7cm
{\Large \bf {Gauge-invariant observables
and marginal deformations}}\\
\vskip 0.4cm {\Large \bf {in open string field theory}} \vskip 1.0cm {\large {Mat\v{e}j
Kudrna,${}^{1}$ Toru Masuda,${}^2$ Yuji Okawa,${}^2$}} \vskip 0.3cm {\large {Martin Schnabl${}^1$
and Kenichiro Yoshida${}^3$}} \vskip 0.8cm
${}^1${\it {Institute of Physics AS CR}}\\
{\it {Na Slovance 2, Prague 8, Czech Republic}}\\
\vskip 0.3cm
${}^2${\it {Institute of Physics, The University of Tokyo}}\\
{\it {3-8-1 Komaba, Meguro-ku, Tokyo 153-8902, Japan}}\\
\vskip 0.3cm
${}^3${\it {Ministry of Internal Affairs and Communications}}\\
{\it {2-1-2 Kasumigaseki, Chiyoda-ku, Tokyo 100-8926, Japan}}\\
\vskip 0.3cm
matej.kudrna@email.cz, masudatoru@gmail.com, okawa@hep1.c.u-tokyo.ac.jp,\\
schnabl.martin@gmail.com, ken16.yoshida@gmail.com

\vskip 1.0cm

{\bf Abstract}
\end{center}

\noindent The level-truncation analysis of open string field theory for a class of periodic
marginal deformations indicates that a branch of solutions in Siegel gauge exists only for a finite
range of values of the marginal field. The periodicity in the deformation parameter is thus
obscure. We use the relation between gauge-invariant observables and the closed string tadpole on a
disk conjectured by Ellwood to construct a map between the deformation parameter of the boundary
conformal field theory and the parameter labeling classical solutions of open string field theory.
We evaluate the gauge-invariant observables for the numerical solutions in Siegel gauge up to level 12
and find that our results qualitatively agree with the analysis by Sen using the energy-momentum
tensor and are consistent with the picture that the finite range of the branch covers one
fundamental domain of the periodic moduli space.
\end{titlepage}

\newpage


\section{Introduction}
\setcounter{equation}{0}

String field theory can potentially be
a background-independent formulation of string theory.
The current construction of string field theory, however,
requires a choice of a consistent background,
and different backgrounds are expected to be described
by classical solutions of the theory
based on the original background we chose.

Thus the first step to address the problem of background independence in string field theory is the
construction of classical solutions. There has been remarkable progress in constructing analytic
solutions in open string field theory~\cite{Witten:1985cc} since the first construction of an
analytic solution for tachyon condensation by one of the authors~\cite{Schnabl:2005gv}. In
particular, analytic solutions for marginal deformations of boundary conformal field theory have
been constructed and intensively studied~\cite{Schnabl:2007az}--\cite{Noumi:2011kn}. A systematic
procedure to construct analytic solutions for marginal deformations to all orders in the
deformation parameter has been presented in~\cite{Kiermaier:2007vu,Kiermaier:2007ki}. On the other
hand, there has been a puzzle associated with marginal deformations in the level-truncation
analysis in Siegel gauge carried out in~\cite{Sen:2000hx}, where the deformation parameter was
treated nonperturbatively. This is the problem we discuss in this paper. Let us first explain the
setup and describe the puzzle.

When open bosonic string theory on a D25-brane is compactified on a circle of the self-dual radius
$\sqrt{\alpha'}$, there are three marginal operators in the boundary conformal field theory (CFT)
describing the compactified direction. The three operators are given by
\begin{equation}
V_1 (t) \equiv
\sqrt{2} : \cos \frac{X (t)}{\sqrt{\alpha'}} : \,, \quad
V_2 (t) \equiv
\sqrt{2} : \sin \frac{X (t)}{\sqrt{\alpha'}} : \,, \quad
V_3 (t) \equiv
\frac{i}{\sqrt{2 \alpha'}} \, \partial_t X (t) \,,
\end{equation}
where $X (t)$ is the coordinate of the compactified direction. The marginal deformation by the
operator $V_3$ corresponds to turning on a constant mode of the gauge field on the D-brane and is
exactly marginal. Since the direction of the coordinate $X$ is compactified, the one-dimensional
moduli space of this deformation is periodic. These three operators are related by the enhanced
$SU(2)$ symmetry at the self-dual radius, and so the marginal deformations by the other two
operators $V_1$ and $V_2$ are also exactly marginal and the moduli space for each deformation is
periodic. It is known that the deformation by the operator $V_1$ changes the original Neumann
boundary condition to a Dirichlet boundary condition at a special point of the moduli space. In
other words, the original D25-brane is deformed to a D24-brane by this marginal deformation.

We thus expect that the equations of motion of open string field theory formulated around a
D25-brane compactified on the self-dual radius have a one-parameter family of solutions associated
with each of the three marginal deformations, and the moduli space is periodic for each case.
In~\cite{Sen:2000hx} Sen and Zwiebach constructed solutions using level truncation in Siegel gauge.
They computed the effective potential of the massless mode associated with the marginal deformation
by solving the equations of motion for other fields. They found that the effective potential is
approximately flat and becomes flatter as the level of the approximation is increased, which is in
accord with the expectation that we have a one-parameter family of solutions. However, they also
found that the branch of the effective potential is truncated at a finite distance from the origin.
The effective potential does not exist beyond that point, and this result seems stable as the
truncation level is increased. It has been a long-standing problem to understand the nature of this
phenomenon. First of all, the periodicity in the moduli space is obscure. In particular, it is
important to know whether or not the point of the moduli space corresponding to a Dirichlet
boundary condition in the case of the deformation by $V_1$ is within the branch.

In order to investigate this problem it is helpful to construct a map between the deformation
parameter $\lambda_{\rm BCFT}$ of boundary CFT and the parameter $\lambda_{\rm SFT}$ labeling
solutions of open string field theory. However, this has in general been a difficult problem.
In~\cite{Sen:2000hx} Sen and Zwiebach attempted to obtain information on the map in the case of the
deformation associated with $V_1$ by slightly increasing the radius $R$ of the compactification.
The marginal deformation becomes a relevant deformation and the effective potential develops a
local minimum corresponding to tachyon condensation. They used the location of the local minimum to
identify the point of the moduli space corresponding to a Dirichlet boundary condition. However,
the extrapolation to the self-dual radius $R \to \sqrt{\alpha'}$ is not smooth, and they were not
able to obtain a definite conclusion.

Later Sen developed a different method to construct a map between $\lambda_{\rm BCFT}$ and
$\lambda_{\rm SFT}$ in~\cite{Sen:2004cq}. It is based on the energy-momentum tensor in spacetime.
Its dependence on $\lambda_{\rm BCFT}$ can be calculated from the boundary state. Its dependence on
$\lambda_{\rm SFT}$ was calculated from the dependence of the effective potential on the
compactification radius $R$. The two results were combined and a map between $\lambda_{\rm BCFT}$
and $\lambda_{\rm SFT}$ was constructed based on numerical results by level truncation up to
level~4.\footnote{
The analysis was extended to higher levels by Andr\'{e} Kurs
in his Senior Thesis at Princeton University~\cite{Kurs}.}
In order to use this method, however, it is necessary to calculate the dependence of the
effective potential on the compactification radius $R$. Furthermore, this method cannot be used for
arbitrary marginal deformations. For instance, it cannot be used directly for the marginal
deformation by $V_3$ because the relevant component of the energy-momentum tensor does not depend
on the deformation parameter, although we can use the $SU(2)$ symmetry to convert the result for
$V_1$ into that for $V_3$.

In this paper we present a new approach to the construction of a map between the deformation
parameter $\lambda_{\rm BCFT}$ of boundary CFT and the parameter $\lambda_{\rm SFT}$ labeling
solutions of open string field theory for marginal deformations. It is based on a relation between
gauge-invariant observables discovered in~\cite{Hashimoto:2001sm, Gaiotto:2001ji} and the closed
string tadpole on a disk conjectured by Ellwood~\cite{Ellwood:2008jh}. If the closed string tadpole
depends on the marginal deformation of boundary CFT, we can construct a map between $\lambda_{\rm
BCFT}$ and $\lambda_{\rm SFT}$ by calculating the gauge-invariant observable for the solutions of
open string field theory. This method can in principle be used for any marginal deformation if
there is a closed string tadpole which depends on the deformation parameter. Furthermore, the
calculation on the string field theory side is much simpler than that of the method
in~\cite{Sen:2004cq}. The gauge-invariant observables depend linearly on the open string field and
can be calculated by contracting the open string field and the identity state with an additional
insertion of an on-shell closed string vertex operator.

The paper is organized as follows. In section~\ref{s-Marginal} we briefly review the construction
of the solutions for marginal deformations by Sen and Zwiebach in level truncation. In
section~\ref{s-Ellwood} we apply the conjecture by Ellwood to the solutions for marginal
deformations and explain how one can relate the parameter of the solutions in open string field
theory with that of the corresponding boundary CFT. In section~\ref{s-level4} we illustrate our
computation using level truncation up to level~4. In section~\ref{s-level12} we summarize our main
numerical results for solutions with various values of $\lambda_{\mathrm{SFT}}$ obtained up to
level~12. Our results are consistent with the picture that the finite range of the branch covers
just one fundamental domain of the periodic moduli space. Further supporting tables and plots are
presented in appendix~\ref{a-tables}.

\section{Marginal deformations in Siegel gauge}
\label{s-Marginal}
\setcounter{equation}{0}

In this section we briefly review the result by Sen and Zwiebach in~\cite{Sen:2000hx} for marginal
deformations in open string field theory using level truncation. We consider open string field
theory for a D25-brane in a 26-dimensional flat spacetime with one of the spatial directions
compactified on a circle of the self-dual radius and the marginal deformation generated by $V_1
(t)$.

The important features can already be seen at level~1.
The string field truncated to level~1 is given by
\begin{equation}
t_0 \, c_1 \ket{0}
+ t_1 : \cos \frac{X(0)}{\sqrt{\alpha'}} : c_1 \ket{0} \,,
\end{equation}
and the potential for the modes $t_0$ and $t_1$
with the normalization used in~\cite{Sen:2000hx} is
\begin{equation}
V (t_0, t_1) = - \frac{1}{2} \, t_0^2
+ \frac{27 \sqrt{3}}{64} \, t_0^3
+ \frac{3 \sqrt{3}}{8} t_0 \, t_1^2 \,.
\end{equation}
Let us derive the effective potential for $t_1$
by solving the equation of motion for $t_0$.
Since the equation of motion for $t_0$ is a quadratic equation,
there are two solutions:
\begin{equation}
t_0^M = \frac{4}{81 \sqrt{3}}
\left( - \sqrt{64 - \frac{729}{2} \, t_1^2} + 8 \right) \,, \qquad
t_0^V = \frac{4}{81 \sqrt{3}}
\left( \sqrt{64 - \frac{729}{2} \, t_1^2} + 8 \right) \,.
\end{equation}
The superscript $M$ denotes the {\it marginal} branch.
The solution $t_0^M$ vanishes as $t_1 \to 0$
so that this branch is associated with the marginal deformation.
The superscript $V$ denotes the {\it vacuum} branch.
The solution $t_0^V$ is associated
with the branch connected to the tachyon vacuum solution.
We can see from these expressions that
we do not have real solutions when
\begin{equation}
64 - \frac{729}{2} \, t_1^2 < 0 \,.
\end{equation}
The critical values $\pm \bar{t}_1$ for $t_1$ are thus given by
\begin{equation}
\label{lambda_crit}
\bar{t}_1 = \frac{8 \sqrt{2}}{27}
\simeq 0.419 \,.
\end{equation}
The two branches meet at these critical values.
Since the effective potential is even with respect to $t_1$,
from now on we focus our attention on the region $t_1 \ge 0$.

\begin{figure}[h]
\centerline{\hbox{\epsfig{figure=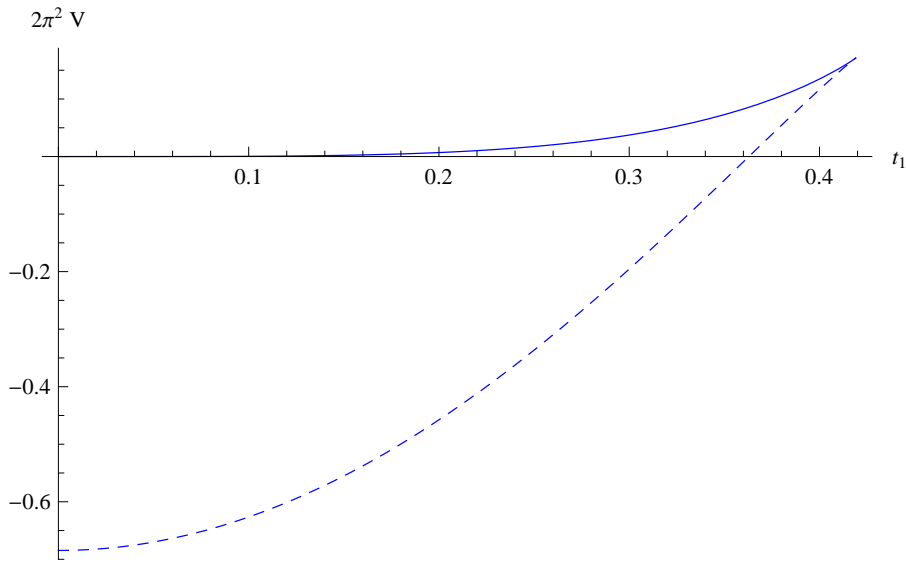, width=12cm}}}
\caption{The two branches of the effective potential for $t_1$
at level~1.
The solid line is for $2 \pi^2 V^M (t_1)$
and the dashed line is for $2 \pi^2 V^V (t_1)$.}
\label{figure-1}
\end{figure}

Corresponding to the two solutions $t_0^M$ and $t_0^V$,
there are two branches for the effective potential.
By substituting $t_0^M$ and $t_0^V$ to $V(t_0, t_1)$, we obtain
\begin{equation}
\begin{split}
V^M (t_1) & = \frac{2}{59049} \,
\Bigl( \, -512 + 4374 \, t_1^2
+ ( \, 64 -\frac{729}{2} \, t_1^2 \, )^{\frac{3}{2}} \, \Bigr)
= \frac{27}{128} \, t_1^4 + \frac{6561}{32768} \, t_1^6
+ {\cal O}(t_1^8) \,, \\
V^V (t_1) & = \frac{2}{59049} \,
\Bigl( \, -512 + 4374 \, t_1^2
- ( \, 64 -\frac{729}{2} \, t_1^2 \, )^{\frac{3}{2}} \, \Bigr)
= - \frac{2048}{59049} + \frac{8}{27} \, t_1^2
+ {\cal O}(t_1^4) \,.
\end{split}
\end{equation}
The two branches $V^M (t_1)$ and $V^V (t_1)$
of the effective potential are shown in figure~\ref{figure-1}.

\begin{figure}[ht]
 \begin{minipage}[b]{0.5\linewidth}
 \centering
 \resizebox{3.2in}{1.7in}{\includegraphics[scale=1]{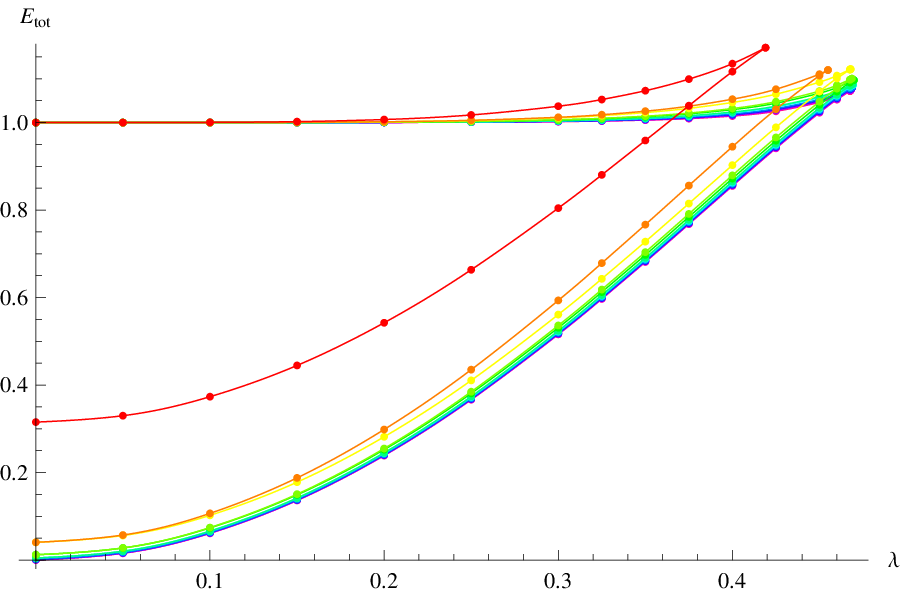}}
 \end{minipage}
 \hspace{-0.2cm}
 \begin{minipage}[b]{0.5\linewidth}
 \centering
 \resizebox{3.2in}{1.7in}{\includegraphics[scale=1]{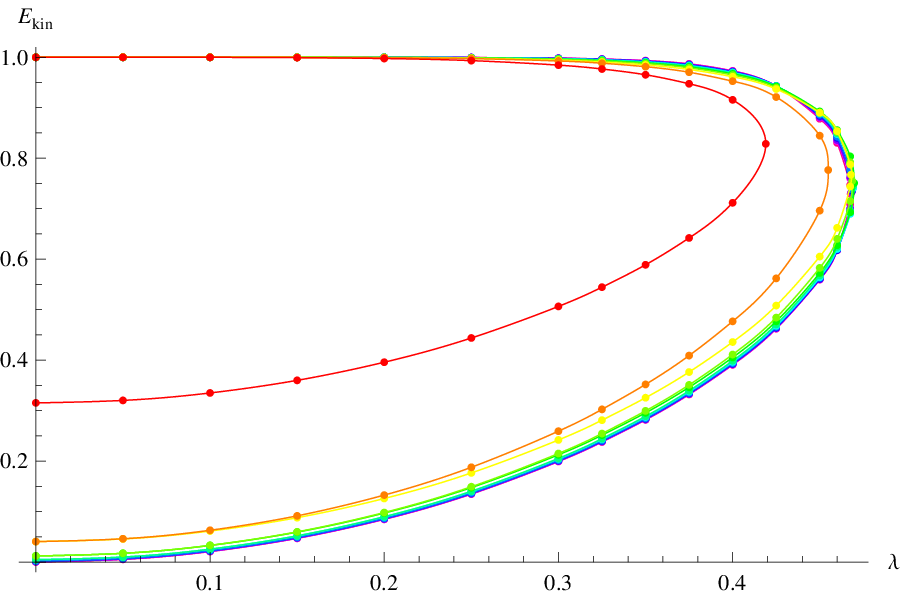}}
 \end{minipage}
 \caption{\label{fig:energy}{The energy density
 of the marginal and vacuum branches computed from the full action (the left graph)
 and from the kinetic term (the right graph).
 We have chosen $t_1$ as the parameter $\lambda$ of the branches.
 The energy density is measured from the tachyon vacuum
and normalized by the D25-brane tension.
 In this and other figures, the colors follow the spectrum: higher level results are depicted using shorter wavelength colors.
 For explicit color coding, see figure~\ref{Fig:legend}.}}
 \end{figure}

We expect an exactly flat potential for the marginal branch because the deformation by $V_1 (t)$ is
known to be exactly marginal. The potential $V^M (t_1)$ at level~1 is not exactly flat, but this is
considered to be an artifact of level truncation. The higher-order analysis in level truncation
shows that the effective potential for $t_1$ on the marginal branch becomes flatter as the
truncation level is increased, as expected. See figure~\ref{fig:energy}.\footnote{The difference
between the energy density $E_{tot}$ computed from the full action and the energy density $E_{kin}$
computed from the kinetic term is proportional to
$t_1$ times the equation of motion for $t_1$. }
It has also been shown
analytically that the coefficient in front of $t_1^4$ in $V^M (t_1)$ vanishes in the limit where
the truncation level becomes infinite~\cite{Berkovits:2003ny}. In this deformation, we can
therefore use $t_1$ as the label $\lambda_{\rm SFT}$ of approximate numerical solutions in level
truncation.

On the other hand, the existence of the critical value of $t_1$ does not seem to be an artifact of
level truncation. The results at higher orders in figure~\ref{fig:energy} show that the critical
value persists and the position of the critical value does not move significantly as the truncation
level is increased. The analysis using level truncation thus indicates that the effective potential
$V^M (t_1)$ on the marginal branch in Siegel gauge becomes exactly flat and terminates at a finite
critical value in the limit where the truncation level becomes infinite.

\section{Gauge-invariant observables and the closed string tadpole}
\setcounter{equation}{0}
\label{s-Ellwood}

In~\cite{Ellwood:2008jh} Ellwood conjectured a relation between gauge-invariant observables of open
string field theory discovered in~\cite{Hashimoto:2001sm, Gaiotto:2001ji} and the closed string
tadpole on a disk. The gauge-invariant observable $W ( \phi \,, \mathcal{V} )$ for an open string
state $\phi$ and an on-shell closed string vertex operator $\mathcal{V}$ is defined by the
following correlation function on the upper half-plane:
\begin{equation}
W ( \phi \,, \mathcal{V} )
= \langle \, \mathcal{V} (i) \, f_I \circ \phi (0) \,
\rangle_{\rm UHP} \,.
\end{equation}
Here $\phi (0)$ is the operator corresponding to the state $\phi$
in the state-operator correspondence
and we denoted by $f_I \circ \phi (0)$
the conformal transformation of $\phi (0)$ under the map $f_I (\xi)$ associated with the
identity state:
\begin{equation}
f_I (\xi) = \frac{2 \, \xi}{1 - \xi^2} \,.
\end{equation}
The closed string tadpole on a unit disk
defined in~\cite{Ellwood:2008jh} is
\begin{equation}
\label{tadpole-disk}
\mathcal{A}^{\rm disk} ( \mathcal{V} )
= -\frac{e^{-i \theta}}{2 \pi i} \,
\langle \, \mathcal{V} (0) \, c (e^{i \theta}) \,
\rangle_{\rm disk} \,.
\end{equation}
This should be independent of $\theta$.
The unit disk with a complex coordinate $w$ can be mapped
to the upper half-plane of $z$
by the following conformal transformation:
\begin{equation}
z = i \, \frac{1-w}{1+w} \,.
\end{equation}
The correlation function on a unit disk in~(\ref{tadpole-disk})
can be mapped under this conformal transformation to
\begin{equation}
\mathcal{A}^{\rm disk} ( \mathcal{V} )
= - \frac{1}{\pi} \,
\frac{1}{1 + t^2} \,
\langle \, \mathcal{V} (i) \,
c (t) \, \rangle_{\rm UHP}
\quad \text{with} \quad
t = \tan \frac{\theta}{2} \,.
\end{equation}
It is easy to confirm that
$\mathcal{A}^{\rm disk} ( \mathcal{V} )$
is independent of $t$
when the ghost part of $\mathcal{V}$ is $c \tilde{c}$.
Ellwood conjectured the following relation:
\begin{equation}
W ( \Psi, \mathcal{V} )
= \mathcal{A}^{\rm disk}_\Psi ( \mathcal{V} )
- \mathcal{A}^{\rm disk}_0 ( \mathcal{V} ) \,,
\label{Ellwood-conjecture}
\end{equation}
where $\mathcal{A}^{\rm disk}_0$ is the closed string tadpole
with the original boundary condition
and $\mathcal{A}^{\rm disk}_\Psi$ is the closed string tadpole
with the boundary condition
corresponding to the classical solution $\Psi$.
If the boundary CFT has marginal deformations
labeled by $\lambda_{\rm BCFT}$,
we expect that the equation of motion of open string field theory
has a one-parameter family of solutions
labeled by $\lambda_{\rm SFT}$.
The left-hand side of~(\ref{Ellwood-conjecture})
is a function of $\lambda_{\rm SFT}$.
The closed string tadpole appearing
on the right-hand side of~(\ref{Ellwood-conjecture})
is a function of $\lambda_{\rm BCFT}$.
We can thus obtain a map
between $\lambda_{\rm BCFT}$ and $\lambda_{\rm SFT}$ from
the conjectured relation~(\ref{Ellwood-conjecture}).

We consider the marginal deformation by the cosine potential $V_1 (t)$. We have to choose a closed
string vertex operator $\mathcal{V}$ such that the one-point function $\mathcal{A}^{\rm disk} (
\mathcal{V} )$ has a nontrivial dependence on $\lambda_{\rm BCFT}$. We choose
\begin{equation}
\label{cal-V}
\mathcal{V} = - \frac{2i}{\alpha'} \,
c \tilde{c} \partial X \bar{\partial} X \,.
\end{equation}
The normalization of $\mathcal{V}$ is the same as
that in~\cite{Kawano:2008ry}.
The dependence of $\mathcal{A}^{\rm disk} ( \mathcal{V} )$ on $\lambda_{\rm BCFT}$
can be easily determined using the boundary state~\cite{Recknagel:1998ih},
as studied, for example, in~\cite{Sen:2002nu}.
We have
\begin{equation}
\mathcal{A}^{\rm disk} ( \mathcal{V} )
\propto \cos \, ( \, 2 \pi \lambda_{\rm BCFT} \, ) \,.
\end{equation}
It is a periodic function of $\lambda_{\rm BCFT}$,
and the Dirichlet boundary condition corresponds to
the point $\lambda_{\rm BCFT} = 1/2$.
The overall normalization of $\mathcal{A}^{\rm disk} ( \mathcal{V} )$
can be determined by evaluating the one-point function
of $\mathcal{V}$ on a disk with the Neumann boundary condition,
which corresponds to $\lambda_{\rm BCFT} = 0$.
Our normalization of correlation functions on the upper half-plane is
\begin{equation}
\langle \, c(z_1) \, c(z_2) \, c(z_3) \, \rangle_{\rm UHP}
= (z_1 - z_2) (z_1 - z_3) (z_2 - z_3) \,.
\end{equation}
Here and in what follows
we divide correlation functions
by the spacetime volume factor.
The one-point function of $\mathcal{V}$
on the upper half-plane with the Neumann boundary condition is given by
\begin{equation}
-\frac{1}{\pi} \,
\frac{1}{1 + t^2} \,
\Bigl( \, -\frac{2i}{\alpha'} \, \Bigr)
\langle \, c \tilde{c} \partial X \bar{\partial} X (i) \,
c (t) \, \rangle_{\rm UHP}
= -\frac{1}{2 \pi} \,.
\end{equation}
This fixes the overall normalization of $\mathcal{A}^{\rm disk} ( \mathcal{V} )$
to give
\begin{equation}
\mathcal{A}^{\rm disk} ( \mathcal{V} )
= -\frac{1}{2 \pi} \,
\cos \, ( \, 2 \pi \lambda_{\rm BCFT} \, ) \,.
\end{equation}
Using the relation~(\ref{Ellwood-conjecture}) we have
\begin{equation}
\label{W-lambda_BCFT}
W ( \Psi, \mathcal{V} )
= -\frac{1}{2 \pi} \, \Bigl[ \,
\cos \, ( \, 2 \pi \lambda_{\rm BCFT} \, ) - 1 \,
\Bigr] \,.
\end{equation}
It is convenient to introduce
\begin{equation}
\label{W_XX}
W_{XX}
\equiv 1 - 2 \pi \, W ( \Psi, \mathcal{V} )
\end{equation}
for $\mathcal{V}$ given by (\ref{cal-V}).
Then we have
\begin{equation}
W_{XX}
= \cos \, ( \, 2 \pi \lambda_{\rm BCFT} \, ) \,.
\label{map}
\end{equation}
As we mentioned before, the left-hand side is
a function of $\lambda_{\rm SFT}$ labeling the solutions so that we can
derive a map between $\lambda_{\rm SFT}$ and $\lambda_{\rm BCFT}$ from this relation. In
particular, the value of the gauge-invariant observable for the solution corresponding to the
Dirichlet boundary condition $\lambda_{\rm BCFT} = 1/2$ is given by $W_{XX} = -1$.

\section{Evaluation of the gauge-invariant observables}
\setcounter{equation}{0}
\label{s-level4}

In this section we illustrate our computation of the gauge-invariant observables
for the solutions corresponding to the cosine deformations reviewed in section~\ref{s-Marginal}.
The solutions in Siegel gauge
were constructed by Sen and Zwiebach in~\cite{Sen:2000hx}
up to level~4.
We expand the string field in level $\ell$ as follows:
\begin{equation}
| \, \Psi \, \rangle
= \sum_{\ell = 0}^\infty | \, \Psi^{(\ell)} \, \rangle \,.
\end{equation}
The expressions up to level~4
in the notation used in section~3 of~\cite{Sen:2000hx}
are given by
\begin{equation}
\begin{split}
\label{psi}
| \, \Psi^{(0)} \, \rangle
& = t_0 \, c_1 \ket{0} \,, \\
| \, \Psi^{(1)} \, \rangle
& = t_1  : \cos \frac{X(0)}{\sqrt{\alpha'}} : c_1 \ket{0} \,, \\
| \, \Psi^{(2)} \, \rangle
& = \bigl( \, u_0\,c_{-1}\, b_{-1} + v_0\, L_{-2}^X
+w_0\,L_{-2}^{\prime}  \, \bigr) \, c_1\, \ket{0} \,, \\
| \, \Psi^{(3)} \, \rangle
& = \bigl( \, u_1\, c_{-1}\, b_{-1} + v_1\, L_{-2}^X +w_1\,L_{-2}^{\prime}
+ z_1 L_{-1}^X \,L_{-1}^X \, \bigr) \,\ket{\varphi_t}  \quad
\text{with} \quad
\ket{\varphi_t} = {}: \cos \frac{X(0)}{\sqrt{\alpha'}} : c_1 \ket{0}\,, \\
| \, \Psi^{(4)} \, \rangle
& = \tilde g \, \ket{p_4} + t_2 \, \ket{\chi}
+\Bigl[ \, a\,L_{-4}^X + \bar a\, L_{-4}^{\prime}
+b\, L_{-2}^X\, L_{-2}^X +\bar b\, L_{-2}^{\prime}\, L_{-2}^{\prime}
+\hat b\, L_{-2}^{\prime}\, L_{-2}^X \\
& \qquad \quad
+ c \, c_{-3}\, b_{-1}
+ d \, b_{-3}\, c_{-1}+ e\, b_{-2}\, c_{-2}
+ ( \, f \, L^X_{-2} +\bar f\, L^{\prime}_{-2} \, ) \, c_{-1}\, b_{-1} \, \Bigr] \,
c_1 \ket{0} \,,
\end{split}
\end{equation}
where
$L_n^{X}$ are the Virasoro generators associated with the field $X (t)$ describing the compactified
direction and $L_{n}^{\prime}$ are the Virasoro generators associated with the rest of the matter
fields. The primary field $\ket{p_4}$ at level~4 is given by
\begin{equation}
\ket{p_4}
=\bigl( \, \alpha_{-3}^X\, \alpha_{-1}^X
-\frac{3}{4}\,(\alpha_{-2}^X)^2
-\frac{1}{2}\,(\alpha_{-1}^X)^4 \, \bigr) \, c_1\, \ket{0} \,,
\end{equation}
and $\ket{\chi}$ is
\begin{equation}
\ket{\chi}= {}: \cos \frac{2\, X(0)}{\sqrt{\alpha'}}  : \ket{0} \,.
\end{equation}

The gauge-invariant observable $W ( \phi, \mathcal{V} )$
can be expressed as a BPZ inner product of $\phi$ with an open string state.
For $\mathcal{V}$ given by (\ref{cal-V}),
the gauge-invariant observable is
\begin{equation}
W ( \phi, \mathcal{V} )
= \langle \, \Phi_{XX} \,, \phi \, \rangle \,,
\end{equation}
where the open string state $\Phi_{XX}$ has been constructed in~\cite{Kawano:2008ry} and is given
by
\begin{equation}
\ket{\Phi_{XX}}
= \left (\frac{1}{4}
-4\sum_{n,\,m=1}^{\infty}
i^{m-n}\,m\,n\,\alpha^X_{-m}\,\alpha _{-n}^X \right )
e^{E}c_0\,c_1\,\ket{0}
\end{equation}
with
\begin{equation}
E=-\frac{1}{2} \sum_{n=1}^\infty \,
\frac{(-1)^n}{n}\alpha_{-n}\cdot \alpha_{-n}
+ \sum _{n=1}^\infty \, (-1)^n\,c_{-n}\,b_{-n}\,.
\end{equation}
We denote the gauge-invariant observable
$W ( \Psi, \mathcal{V} )$ truncated to level~$\ell$
by $W^{(\ell)} ( \Psi, \mathcal{V} )$.
The quantity $W^{(4)} ( \Psi, \mathcal{V} )$
is given by
\begin{equation}
\begin{split}
\label{W^(4)}
W^{(4)} ( \Psi, \mathcal{V} )
&= \sum_{\ell = 0}^4 \,
\langle \, \Phi_{XX}, \Psi^{(\ell)} \, \rangle\\
&= \frac{1}{4} \, t_0
+ \frac{1}{4} \, u_0 - \frac{31}{8} \, v_0 + \frac{25}{8} \, w_0
+ \frac{31}{4} \, a -\frac{25}{4} \, \bar a
+ \frac{963}{16} \, b + \frac{675}{16} \, \bar b
-\frac{775}{16}\, \hat b \\
& \quad~ + \frac{1}{4}\, e
-\frac{31}{8}\, f +\frac{25}{8}\, \bar f +192 \, \tilde g \,.
\end{split}
\end{equation}
Note that there are no contributions from
$\Psi^{(1)}$ and $\Psi^{(3)}$
because of momentum conservation.

In the numerical solution constructed by Sen and Zwiebach in~\cite{Sen:2000hx}, the component
fields $t_0$, $t_1$, $u_0$, $\ldots$ are given as functions of $t_1$. We therefore obtain $W^{(4)}
( \Psi, \mathcal{V} )$ for the solution as a function of $t_1$, which we are using as the label
$\lambda_{\rm SFT}$. This is how we compute the gauge-invariant observable as a function of
$\lambda_{\rm SFT}$ in level truncation. We then compare the result with the
expression~(\ref{W-lambda_BCFT}) of the gauge-invariant observable as a function of $\lambda_{\rm
BCFT}$ to find a relation between $\lambda_{\rm SFT}$ and $\lambda_{\rm BCFT}$ numerically.

\section{Numerical evaluation to level 12}
\setcounter{equation}{0}
\label{s-level12}

The method illustrated in the preceding section at level~4 can be extended to higher levels. We had
generated all required vertices using the conservation laws of \cite{Rastelli:2000iu} and then
solved the equations of motion
by Newton's iterative method.\footnote{
Newton's method requires
a choice of an initial approximate solution, which is then iteratively improved to any desired
accuracy. We define it as $\frac{\|\Psi^{(i)}-\psi^{(i-1)}\|}{\|\psi^{(i)}\|}$ and stop the
iteration when we reach $10^{-8}$.
The  norm is defined, as usual, by a square root of the sum of coefficients squared
in the basis formed by $\alpha$ oscillators of the compact direction,
$bc$ oscillators of the ghost sector,
and the Virasoro generators in the rest of the matter sector. 
As the starting point we either used results for the solution
with the same $\lambda$ found at lower level or, sometimes more efficiently, used a different
solution with a neighboring value of $\lambda$ at the same level.
 } With
the help of a computer cluster we were able to perform all our computations to level~$(12, 36)$.

To avoid possible confusion between the marginal and vacuum branches, but also out of curiosity, we studied numerically both branches for various values of $\lambda$, up to the critical
level-dependent $\lambda_{crit}$, where both branches meet. The value of $\lambda_{crit}$ at level
1 is given by the simple analytic formula (\ref{lambda_crit}). At higher levels we employed the
interval bisection method and studied
whether Newton's method
converges for given $\lambda$.
The results are given in table \ref{t-lambda_crit}.
\begin{table}[h]
\begin{equation}
\begin{array}{|c|c|r|c|}\hline
L & \lambda_{crit} &L & \lambda_{crit}\\\hline
1  & 0.419026 & 7  & 0.469239\\
2  & 0.454866 & 8  & 0.468645\\
3  & 0.467900 & 9  & 0.468623\\
4  & 0.468614 & 10 & 0.468160\\
5  & 0.469761 & 11 & 0.468109\\
6  & 0.469100 & 12 & 0.467748\\\hline
\end{array}\nonumber
\end{equation}
\caption{Results for $\lambda_{crit}$ at levels $L=1,\ldots,12$. The values are rounded down to six
digits, so that the solution still exists for these values. It can be found within about 5
iterations if we use as the starting point
for Newton's method
the highest $\lambda$ solution
found in the previous steps at the same level. Increasing the last digit by one,
Newton's method
would not anymore converge within~50 iterations at lower levels and at least~10 iterations
at level~12.} \label{t-lambda_crit}
\end{table}
Interestingly, the critical value grows at low levels and is largest at level 5, and after one
oscillation it monotonically decreases. Simple linear extrapolation in $1/L$ to the infinite level
gives the critical value of about $0.466$.
Values of the energy and the gauge-invariant observables
at $\lambda_{crit}$ for each given level $L=1,\ldots,12$ are presented
in table \ref{t-critical} of appendix~\ref{a-tables}.

\begin{figure}[ht]
 \begin{minipage}[b]{0.5\linewidth}
 \centering
 \resizebox{3.2in}{1.7in}{\includegraphics[scale=1]{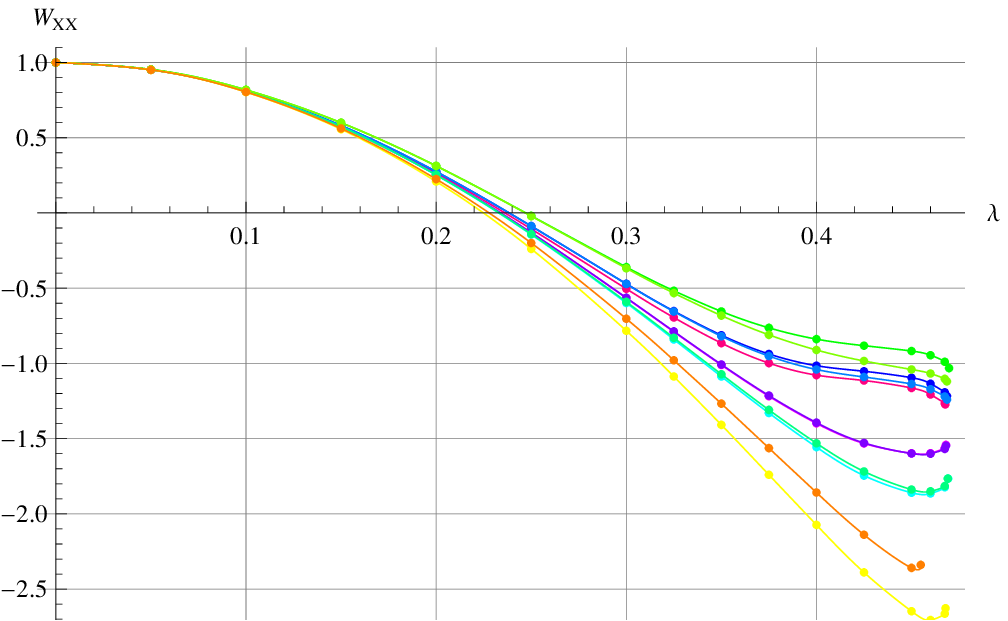}}
 \end{minipage}
 \hspace{-0.2cm}
 \begin{minipage}[b]{0.5\linewidth}
 \centering
 \resizebox{3.2in}{1.7in}{\includegraphics[scale=1]{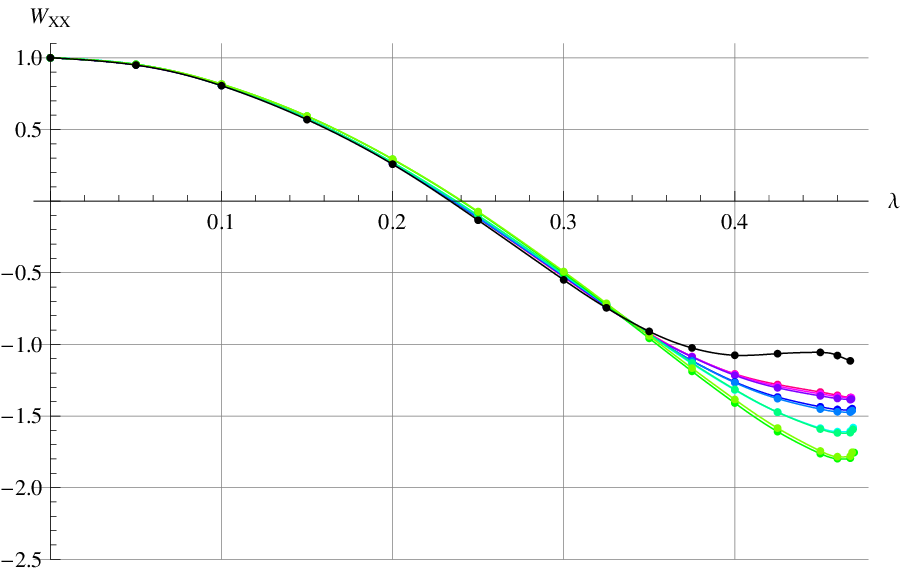}}
 \end{minipage}
 \caption{\label{fig:WXX_M}{
 The values of $W_{XX}$ at levels $L=2,\ldots,12$ (left) and their Pad\'{e}-Borel improvements at levels $L=4,\ldots,12$, together with a fit to the infinite level in black (right).}}
 \end{figure}

We have constructed the marginal and vacuum branches for about 15 different values of $\lambda$.
Figure \ref{fig:energy}, presented in section~\ref{s-Marginal}, shows the energy density of the
marginal and vacuum branches meeting at $\lambda_{crit}$. The quantity $W_{XX}$ defined
in~(\ref{W_XX}) on the marginal branch for various levels is plotted in the left graph of
figure~\ref{fig:WXX_M}. Its dependence on the level is rather erratic, so we tried to apply
Pad\'{e}-Borel resummation of contributions of different levels for each solution. The right graph
of figure~\ref{fig:WXX_M} shows the resulting improved values of $W_{XX}$, together with a simple
extrapolation to the infinite level. More data are presented in appendix \ref{a-tables}.

One may notice that starting at level 3 the results at odd levels are close to the values found at
the previous even level. The reason is most likely that at even levels new fields with zero
momentum appear, and they seem to have larger influence than the fields with nonzero momentum. The
similarity of odd and even levels is even greater for $W_{XX}$ because it receives contributions
only from fields at even levels, meaning that any change at odd levels can arise only from the
change of the solution and not from
adding new fields.

Table \ref{t-observables} summarizes linear extrapolations in $1/L$ of various quantities of
interest to the infinite level
together with the estimated statistical error.
We estimate the error by considering extrapolations
using different numbers of data points
and computing the sample standard
deviation for such results. For $W_{XX}$ we use the Pad\'{e}-Borel resummed version. We also
computed $W_{00}$ defined by
\begin{equation}
W_{00} = 1 +2 \pi \, W ( \Psi, \mathcal{V} )
\end{equation}
with
\begin{equation}
\mathcal{V} = - \frac{2i}{\alpha'} \,
c \tilde{c} \partial X^0 \bar{\partial} X^0 \,.
\end{equation}
Up to approximately $\lambda=0.325$ the extrapolated values of the energy density $E_{tot}$
computed from the full action, the energy density $E_{kin}$ computed from the kinetic term, and
$W_{00}$ are consistent with the expected value 1. Apparently, for higher values of $\lambda$ our
method underestimates the error of the extrapolated values.

\begin{table}
\begin{equation}
\begin{array}{|l|ll|ll|ll|ll|}\hline
 \lambda & E_{tot} & \sigma_{E_{tot}} & E_{kin} & \sigma_{E_{kin}} & W_{00} &\sigma_{W00} & W_{XX} & \sigma_{W_{XX}}\\ \hline
 0.05 & 1. & 6.3 \times 10^{-7} & 1. & 2.1 \times 10^{-7} & 1. & 0.00007 & 0.9493 & 0.0018 \\
 0.1 & 1. & 0.00001 & 1. & 3.4 \times 10^{-6} & 1. & 0.00028 & 0.8051 & 0.0025 \\
 0.15 & 1. & 0.000051 & 1. & 0.000018 & 1.0001 & 0.00066 & 0.5686 & 0.0046 \\
 0.2 & 1. & 0.00016 & 1. & 0.000057 & 1. & 0.0012 & 0.2572 & 0.014 \\
 0.25 & 0.9999 & 0.0004 & 1. & 0.00014 & 0.9996 & 0.002 & -0.1338 & 0.012 \\
 0.3 & 0.9999 & 0.00083 & 0.9997 & 0.00026 & 0.9978 & 0.0029 & -0.5499 & 0.0022 \\
 0.325 & 1. & 0.0011 & 0.999 & 0.00033 & 0.9954 & 0.0034 & -0.7449 & 0.01 \\
 0.35 & 1.0004 & 0.0015 & 0.9972 & 0.00041 & 0.9905 & 0.0039 & -0.9096 & 0.013 \\
 0.375 & 1.0016 & 0.0019 & 0.992 & 0.00052 & 0.9798 & 0.0043 & -1.0251 & 0.027 \\
 0.4 & 1.0049 & 0.0023 & 0.9781 & 0.00035 & 0.9565 & 0.0046 & -1.076 & 0.044 \\
 0.425 & 1.013 & 0.0028 & 0.9433 & 0.0013 & 0.9075 & 0.005 & -1.0649 & 0.059 \\
 0.45 & 1.0321 & 0.0035 & 0.8668 & 0.0046 & 0.8116 & 0.0064 & -1.0562 & 0.076 \\
 0.46 & 1.0451 & 0.0037 & 0.8112 & 0.0067 & 0.7453 & 0.0077 & -1.0778 & 0.085 \\
 0.4675 & 1.0585 & 0.0039 & 0.7124 & 0.019 & 0.6326 & 0.02 & -1.1149 & 0.092 \\\hline
\end{array}\nonumber
\end{equation}
\caption{Extrapolation to the infinite level
and its possible statistical error for the four basic
quantities computed at several values of $\lambda$ on the marginal branch. Up to about
$\lambda=0.325$ the first three observables are within the $3\sigma$ tolerance away from the
expected value of $1$.} \label{t-observables}
\end{table}

\begin{figure}[ht]
 \centering
 \includegraphics[scale=1.4]{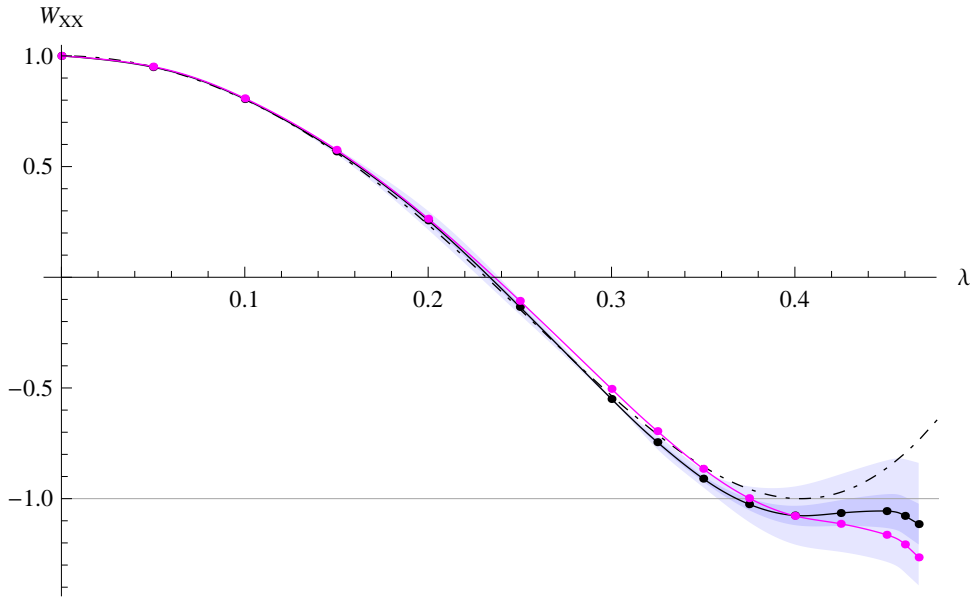}
 \caption{\label{fig:conjecture}{Extrapolation to the infinite level
 of the Pad\'{e}-Borel resummed version of $W_{XX}$.
 The black solid line is the infinite-level fit itself, while the
 shaded regions around it indicate $\sigma$ and $3\sigma$
 uncertainty ranges
 in the extrapolation.
 The magenta line shows the (unimproved) data at level~12 for comparison.
 The black dot-dashed line is
the one-parameter fit of the form $\cos\left(2\pi(\lambda + a \lambda^3)\right)$.}}
 \end{figure}

The results for $W_{XX}$ together with its estimated error are presented in~figure \ref{fig:conjecture}.
The one-parameter fit of the form $\cos\left(2\pi(\lambda + a \lambda^3)\right)$ is also shown,
where the constant $a$ has been fitted
so that  the extrapolation to the infinite level
coincides with the
boundary CFT result at lower values of $\lambda$. The best fit value for $a$ is about $0.99$ at
level 4 and grows at higher levels.
For the extrapolation to the infinite level
we find $a \approx 1.48$.
It would be interesting to calculate
this value analytically. The results of $W_{XX}$ qualitatively
agree with the analysis by Sen~\cite{Sen:2004cq} using the energy-momentum tensor and are
consistent with the picture that the finite range of the branch covers one fundamental domain of
the periodic moduli space. While our results do not provide sufficient evidence that the branch
covers {\it exactly} one fundamental domain, we can safely exclude the possibility that the branch
covers the fundamental domain many times. We can also exclude the opposite possibility that the
branch covers, say, less than 75\% of the fundamental domain. If the branch indeed covers exactly
one fundamental domain, it would be interesting to understand why it is the case.

\section{Discussion}
\setcounter{equation}{0}

In this paper we presented a method of constructing a map between the deformation parameter
$\lambda_{\rm BCFT}$ of boundary CFT and the parameter $\lambda_{\rm SFT}$ labeling solutions of
open string field theory for marginal deformations. While we applied our method to the specific
problem of covering moduli for solutions in Siegel gauge, the basic idea is universal and we can
use it for other problems. For example, a similar problem for the system of separated D-branes has
recently been discussed in~\cite{Karczmarek:2012pn, Longton:2012ei} and it would be interesting to
compute the gauge-invariant observables in the system.

If we construct the boundary state
from numerical solutions
following the proposal in~\cite{Kiermaier:2008qu},
we will be able to obtain more information
on the boundary CFT.
However, construction of the boundary state
for numerical solutions seems to be challenging.
In particular, the closed string state
proposed in~\cite{Kiermaier:2008qu}
is conjectured to reproduce the boundary state
up to a BRST-exact term,
and such a BRST-exact term would obscure
the boundary state especially for approximate solutions
constructed numerically.\footnote{
After this paper was submitted to arXiv,
another approach to the construction of the boundary state
was proposed in~\cite{Kudrna:2012re},
which is more suitable for the application to numerical solutions.}

Our results are qualitatively consistent
with the analysis by Sen~\cite{Sen:2004cq}
using the energy-momentum tensor.
This is not unexpected because both methods
are based on coupling to an infinitesimal closed string field.
However, the closed string is described
by an unintegrated vertex operator
in the gauge-invariant observables
and also in the boundary state construction in~\cite{Kiermaier:2008qu},
while the infinitesimal change in the closed string background
is related more closely to an integrated vertex operator.
It would be an interesting problem
to convert the unintegrated vertex operator
to the integrated vertex operator
in the framework of the gauge-invariant observables
and the boundary state construction~\cite{Kiermaier:2008qu}.

\bigskip

\noindent
\section*{Acknowledgments}

\medskip

We would like to thank Ted Erler, Mitsuhiro Kato, Michael Kiermaier, Carlo Maccaferri and Barton Zwiebach for helpful discussions and to Masaki Murata for careful reading of the manuscript.
The access to computing and storage facilities owned by parties and
projects contributing to the Czech National Grid Infrastructure
MetaCentrum, provided under the programme ``Projects of Large
Infrastructure for Research, Development, and Innovations"
(LM2010005), and to CERIT-SC computing facilities provided under the
programme Center CERIT Scientific Cloud, part of the Operational
Program Research and Development for Innovations, reg. no. CZ.
1.05/3.2.00/08.0144 is highly appreciated.
The work of Y.O. was supported in part
by Grant-in-Aid for Scientific Research~(B) No.~20340048
and Grant-in-Aid for Scientific Research~(C) No.~24540254
from the Japan Society for the Promotion of Science (JSPS).
The work of M.S. was supported by the EURYI grant GACR EYI/07/E010 from EUROHORC and ESF.
The work of Y.O. and M.S. was also supported in part
by the M\v{S}MT contract No. LH11106 and
by JSPS and the
Academy of Sciences of the Czech Republic (ASCR)
under the Research Cooperative Program between Japan and the Czech Republic.

\appendix

\newpage
\section{Tables of numerical results}
\setcounter{equation}{0}
\label{a-tables}

{\normalsize In the following figures $E_{tot}$ denotes the energy density computed from the full
action and $E_{kin}$ denotes the energy density computed from the kinetic term.}

{\footnotesize{
\begin{displaymath}
\begin{array}{|l|llllllll|}\hline
 L/\lambda& 0 & 0.05 & 0.1 & 0.15 & 0.2 & 0.25 & 0.3 & 0.325 \\ \hline
 1  & 1 & 1.00003 & 1.00042 & 1.00216 & 1.00694 & 1.01739 & 1.03737 & 1.05266 \\
 2  & 1 & 1.00001 & 1.0001  & 1.00056 & 1.0019  & 1.00515 & 1.01209 & 1.01789 \\
 3  & 1 & 1.00001 & 1.0001  & 1.00054 & 1.00179 & 1.00473 & 1.01084 & 1.01585 \\
 4  & 1 & 1.      & 1.00005 & 1.00027 & 1.00095 & 1.00264 & 1.00644 & 1.00976 \\
 5  & 1 & 1.      & 1.00005 & 1.00027 & 1.00091 & 1.00249 & 1.00598 & 1.009 \\
 6  & 1 & 1.      & 1.00003 & 1.00018 & 1.00062 & 1.00175 & 1.00436 & 1.00672 \\
 7  & 1 & 1.      & 1.00003 & 1.00017 & 1.00061 & 1.00168 & 1.00413 & 1.00634 \\
 8  & 1 & 1.      & 1.00002 & 1.00013 & 1.00046 & 1.00131 & 1.0033  & 1.00515 \\
 9  & 1 & 1.      & 1.00002 & 1.00013 & 1.00045 & 1.00127 & 1.00317 & 1.00492 \\
 10 & 1 & 1.      & 1.00002 & 1.0001  & 1.00037 & 1.00104 & 1.00266 & 1.00418 \\
 11 & 1 & 1.      & 1.00002 & 1.0001  & 1.00036 & 1.00102 & 1.00258 & 1.00403 \\
 12 & 1 & 1.      & 1.00002 & 1.00009 & 1.00031 & 1.00087 & 1.00223 & 1.00353 \\\hline
\end{array}
\end{displaymath}{{
{\footnotesize{
\begin{displaymath}
\begin{array}{|l|lllllll|}\hline
 L/\lambda& 0.35 & 0.375 & 0.4 & 0.425 & 0.45 & 0.46 & 0.4675 \\\hline
 1  & 1.07282 & 1.09934 & 1.13463 & -       & -       & -       & - \\
 2  & 1.02604 & 1.03744 & 1.05342 & 1.07621 & 1.11061 & -       & - \\
 3  & 1.02282 & 1.03249 & 1.04595 & 1.06492 & 1.09248 & 1.10738 & 1.12122 \\
 4  & 1.01461 & 1.02171 & 1.03215 & 1.04775 & 1.07183 & 1.08537 & 1.09816 \\
 5  & 1.01342 & 1.01992 & 1.02957 & 1.04421 & 1.06719 & 1.08023 & 1.09249 \\
 6  & 1.01029 & 1.01572 & 1.02413 & 1.03747 & 1.05938 & 1.07219 & 1.08449 \\
 7  & 1.00967 & 1.01478 & 1.02277 & 1.03565 & 1.05721 & 1.06995 & 1.08221 \\
 8  & 1.008   & 1.01249 & 1.01977 & 1.03192 & 1.05302 & 1.06575 & 1.07818 \\
 9  & 1.00763 & 1.01191 & 1.01891 & 1.0308  & 1.05177 & 1.06453 & 1.07701 \\
 10 & 1.00658 & 1.01045 & 1.01696 & 1.02837 & 1.04912 & 1.06193 & 1.0746 \\
 11 & 1.00632 & 1.01005 & 1.01636 & 1.02759 & 1.04831 & 1.06117 & 1.07391 \\
 12 & 1.0056  & 1.00902 & 1.01497 & 1.02587 & 1.04647 & 1.0594  & 1.07233 \\\hline
\end{array}
\end{displaymath}}}
\begin{figure}[h]
 \centering
 \includegraphics[width=10cm]{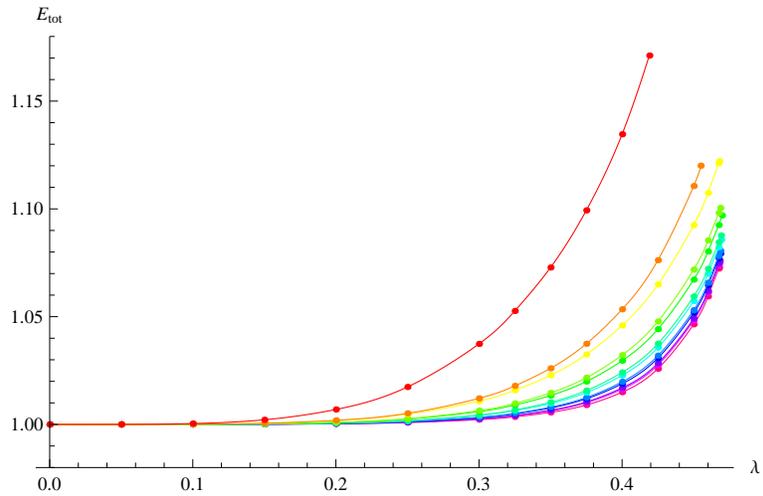}
 \caption{$E_{tot}$ for the marginal branch
 at levels $L=1,\ldots,12$. For the color legend, see figure~\ref{Fig:legend}.}
 \end{figure}

\newpage

{\footnotesize{
\begin{equation}
\begin{array}{|l|llllllll|}\hline
 L/\lambda & 0 & 0.05 & 0.1 & 0.15 & 0.2 & 0.25 & 0.3 & 0.325 \\\hline
 1  & 1 & 0.999991 & 0.999857 & 0.999248 & 0.997483 & 0.993325 & 0.984406 & 0.976723 \\
 2  & 1 & 0.999998 & 0.999962 & 0.999776 & 0.999134 & 0.997311 & 0.992653 & 0.988209 \\
 3  & 1 & 0.999998 & 0.999964 & 0.999796 & 0.999243 & 0.997733 & 0.993941 & 0.990326 \\
 4  & 1 & 0.999999 & 0.999981 & 0.999887 & 0.999541 & 0.998487 & 0.995519 & 0.99246 \\
 5  & 1 & 0.999999 & 0.999982 & 0.999893 & 0.999581 & 0.998645 & 0.996008 & 0.993251 \\
 6  & 1 & 0.999999 & 0.999988 & 0.999925 & 0.999692 & 0.998949 & 0.996713 & 0.994255 \\
 7  & 1 & 0.999999 & 0.999988 & 0.999928 & 0.999711 & 0.999028 & 0.996974 & 0.994692 \\
 8  & 1 & 1.       & 0.999991 & 0.999944 & 0.999769 & 0.999194 & 0.997387 & 0.99531 \\
 9  & 1 & 1.       & 0.999991 & 0.999946 & 0.999779 & 0.999241 & 0.99755  & 0.99559 \\
 10 & 1 & 1.       & 0.999993 & 0.999956 & 0.999815 & 0.999346 & 0.997825 & 0.996017 \\
 11 & 1 & 1.       & 0.999993 & 0.999957 & 0.999821 & 0.999376 & 0.997936 & 0.996215 \\
 12 & 1 & 1.       & 0.999994 & 0.999963 & 0.999845 & 0.999449 & 0.998134 & 0.99653 \\\hline
\end{array}\nonumber
\end{equation}}}

{\footnotesize{
\begin{equation}
\begin{array}{|l|lllllll|}\hline
L/\lambda & 0.35 & 0.375 & 0.4 & 0.425 & 0.45 & 0.46 & 0.4675 \\\hline
1  & 0.965317 & 0.947508 & 0.915633 & -        & -        & -        & - \\
2  & 0.981281 & 0.970361 & 0.952595 & 0.921238 & 0.844668 & -        & - \\
3  & 0.984677 & 0.975766 & 0.961401 & 0.937109 & 0.890234 & 0.853796 & 0.787548 \\
4  & 0.987406 & 0.978993 & 0.964745 & 0.939657 & 0.890215 & 0.852261 & 0.792049 \\
5  & 0.988608 & 0.980684 & 0.966887 & 0.941989 & 0.892573 & 0.855588 & 0.80361 \\
6  & 0.989951 & 0.982294 & 0.968387 & 0.942343 & 0.889422 & 0.849383 & 0.790421 \\
7  & 0.990629 & 0.983238 & 0.969448 & 0.942981 & 0.888695 & 0.847973 & 0.789345 \\
8  & 0.9915   & 0.98433  & 0.970452 & 0.942955 & 0.885679 & 0.8425   & 0.776961 \\
9  & 0.991952 & 0.984971 & 0.971128 & 0.943099 & 0.884412 & 0.84051  & 0.774259 \\
10 & 0.992581 & 0.985794 & 0.971888 & 0.942936 & 0.881779 & 0.835993 & 0.762596 \\
11 & 0.99291  & 0.986275 & 0.972377 & 0.942887 & 0.88049  & 0.834132 & 0.759646 \\
12 & 0.993391 & 0.98693  & 0.972985 & 0.942661 & 0.878228 & 0.830423 & 0.747139 \\\hline
\end{array}\nonumber
\end{equation}}}

\begin{figure}[h]
 \centering
 \includegraphics[width=11cm]{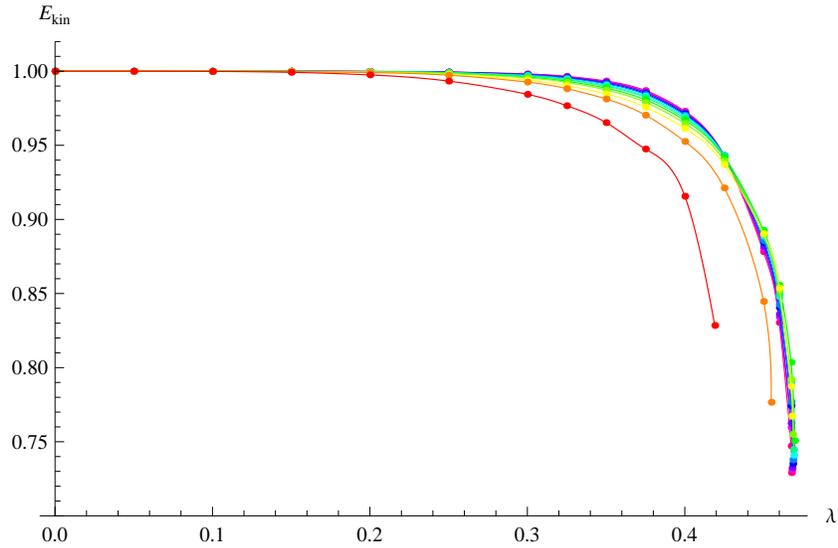}
 \caption{$E_{kin}$ for the marginal branch at levels $L=1,\ldots,12$.}
\end{figure}

\newpage

{\footnotesize{
\begin{equation}
\begin{array}{|l|llllllll|}\hline
 L/\lambda & 0 & 0.05 & 0.1 & 0.15 & 0.2 & 0.25 & 0.3 & 0.325 \\\hline
 1  & 1 & 0.99744  & 0.989648 & 0.976257 & 0.956556 & 0.929248 & 0.891855 & 0.867871 \\
 2  & 1 & 0.999327 & 0.99716  & 0.993028 & 0.986042 & 0.974681 & 0.956329 & 0.94316  \\
 3  & 1 & 0.999329 & 0.997202 & 0.993246 & 0.986732 & 0.976369 & 0.95985  & 0.94805  \\
 4  & 1 & 0.99966  & 0.998549 & 0.996363 & 0.99248  & 0.98572  & 0.973796 & 0.964585 \\
 5  & 1 & 0.999661 & 0.998565 & 0.996445 & 0.992747 & 0.986396 & 0.975226 & 0.966539 \\
 6  & 1 & 0.999773 & 0.999025 & 0.997524 & 0.994784 & 0.989816 & 0.980526 & 0.972937 \\
 7  & 1 & 0.999774 & 0.999033 & 0.997567 & 0.994926 & 0.990191 & 0.981362 & 0.974112 \\
 8  & 1 & 0.999835 & 0.999286 & 0.998173 & 0.9961   & 0.992236 & 0.984701 & 0.978279 \\
 9  & 1 & 0.999835 & 0.999291 & 0.998197 & 0.996186 & 0.99247  & 0.985248 & 0.97907  \\
 10 & 1 & 0.999867 & 0.999423 & 0.998518 & 0.996816 & 0.993596 & 0.987151 & 0.981498 \\
 11 & 1 & 0.999867 & 0.999426 & 0.998534 & 0.996874 & 0.993756 & 0.987539 & 0.982073 \\
 12 & 1 & 0.999891 & 0.999528 & 0.998781 & 0.997362 & 0.994636 & 0.989057 & 0.98404  \\\hline
\end{array}\nonumber
\end{equation}}}

{\footnotesize{
\begin{equation}
\begin{array}{|l|lllllll|}\hline
 L/\lambda &0.35 & 0.375 & 0.4 & 0.425 & 0.45 & 0.46 & 0.4675 \\\hline
 1  & 0.838716 & 0.801585 & 0.748454 & -        & -        & -        & - \\
 2  & 0.926096 & 0.90354  & 0.872612 & 0.826695 & 0.735129 & -        & - \\
 3  & 0.932795 & 0.912723 & 0.885576 & 0.846949 & 0.784627 & 0.74208  & 0.672205 \\
 4  & 0.951971 & 0.934369 & 0.909122 & 0.871148 & 0.807038 & 0.762761 & 0.697553 \\
 5  & 0.954519 & 0.937497 & 0.91265  & 0.874683 & 0.81039  & 0.767018 & 0.709998 \\
 6  & 0.962023 & 0.945903 & 0.921317 & 0.882148 & 0.813582 & 0.766378 & 0.701007 \\
 7  & 0.963589 & 0.947821 & 0.923321 & 0.883583 & 0.813501 & 0.765561 & 0.700451 \\
 8  & 0.96866  & 0.953717 & 0.929562 & 0.888897 & 0.815361 & 0.764379 & 0.691122 \\
 9  & 0.969747 & 0.955078 & 0.930943 & 0.889626 & 0.814543 & 0.762767 & 0.688646 \\
 10 & 0.97277  & 0.958646 & 0.934636 & 0.89228  & 0.814131 & 0.759898 & 0.677343 \\
 11 & 0.973583 & 0.959691 & 0.935684 & 0.892684 & 0.813174 & 0.758292 & 0.674445 \\
 12 & 0.976086 & 0.96273  & 0.938921 & 0.895071 & 0.813104 & 0.756321 & 0.662292 \\\hline
\end{array}\nonumber
\end{equation}}}

\begin{figure}[ht]
 \centering
 \includegraphics[width=11cm]{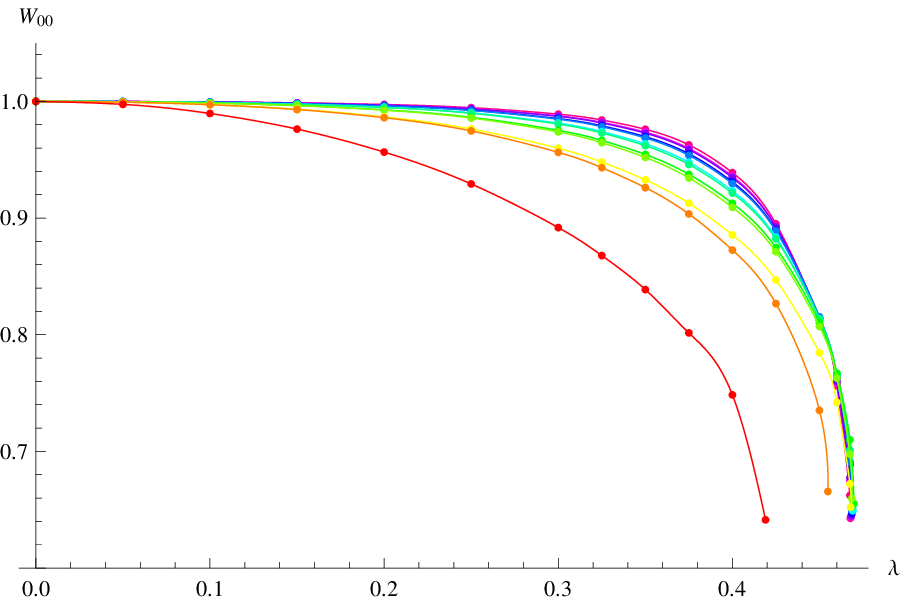}
 \caption{$W_{00}$ for the marginal branch at levels $L=1,\ldots,12$.}
 \end{figure}

\newpage

{\footnotesize{
\begin{equation}
\begin{array}{|l|llllllll|}\hline
 L/\lambda& 0 & 0.05 & 0.1 & 0.15 & 0.2 & 0.25 & 0.3 & 0.325 \\\hline
 1  & 1 & 0.99744  & 0.989648 & 0.976257 & 0.956556 & 0.929248  & 0.891855  & 0.867871  \\
 2  & 1 & 0.951001 & 0.804391 & 0.561486 & 0.225054 & -0.199654 & -0.703015 & -0.979158 \\
 3  & 1 & 0.950938 & 0.803384 & 0.55637  & 0.208866 & -0.238966 & -0.782912 & -1.08704  \\
 4  & 1 & 0.953882 & 0.817906 & 0.599563 & 0.312592 & -0.020983 & -0.368017 & -0.532577 \\
 5  & 1 & 0.953879 & 0.817865 & 0.599431 & 0.312586 & -0.019437 & -0.35992  & -0.516995 \\
 6  & 1 & 0.9517   & 0.808058 & 0.572963 & 0.253324 & -0.140082 & -0.590749 & -0.82979  \\
 7  & 1 & 0.951696 & 0.807988 & 0.572595 & 0.252092 & -0.143343 & -0.598223 & -0.840638 \\
 8  & 1 & 0.952127 & 0.810397 & 0.580849 & 0.274993 & -0.087298 & -0.471833 & -0.655231 \\
 9  & 1 & 0.952126 & 0.810381 & 0.580781 & 0.274838 & -0.087384 & -0.470778 & -0.652329 \\
 10 & 1 & 0.951473 & 0.807389 & 0.572431 & 0.255112 & -0.130855 & -0.563695 & -0.787236 \\
 11 & 1 & 0.951472 & 0.807373 & 0.572346 & 0.254826 & -0.13161  & -0.565422 & -0.789735 \\
 12 & 1 & 0.951627 & 0.808281 & 0.575623 & 0.264426 & -0.106573 & -0.504102 & -0.694964 \\\hline
\end{array}\nonumber
\end{equation}}}
{\footnotesize{
\begin{equation}
\begin{array}{|l|lllllll|}\hline
 L/\lambda & 0.35 & 0.375 & 0.4 & 0.425 & 0.45 & 0.46 & 0.4675 \\\hline
 1  & 0.838716  & 0.801585  & 0.748454  & -        & -        & -        & - \\
 2  & -1.26737  & -1.56275  & -1.85779  & -2.13867 & -2.35874 & -        & - \\
 3  & -1.40816  & -1.74019  & -2.07303  & -2.38872 & -2.64695 & -2.70528 & -2.66319 \\
 4  & -0.682196 & -0.810004 & -0.910679 & -0.98362 & -1.04015 & -1.06733 & -1.10254 \\
 5  & -0.654344 & -0.76347  & -0.838485 & -0.88199 & -0.91762 & -0.94553 & -0.98925 \\
 6  & -1.07165  & -1.30896  & -1.53045  & -1.71789 & -1.83833 & -1.84958 & -1.81515 \\
 7  & -1.08692  & -1.32954  & -1.55612  & -1.74542 & -1.85932 & -1.86422 & -1.823   \\
 8  & -0.819198 & -0.951134 & -1.04046  & -1.09027 & -1.13631 & -1.17153 & -1.2199  \\
 9  & -0.812499 & -0.937299 & -1.01516  & -1.05217 & -1.095   & -1.13498 & -1.19199 \\
 10 & -1.00706  & -1.21363  & -1.39341  & -1.52837 & -1.59857 & -1.60014 & -1.5681  \\
 11 & -1.01053  & -1.21815  & -1.39839  & -1.53177 & -1.59746 & -1.59668 & -1.56202 \\
 12 & -0.865035 & -0.997802 & -1.07794  & -1.11365 & -1.16324 & -1.20662 & -1.26519 \\\hline
\end{array}\nonumber
\end{equation}}}

\begin{figure}[ht]
 \centering
 \includegraphics[width=11cm]{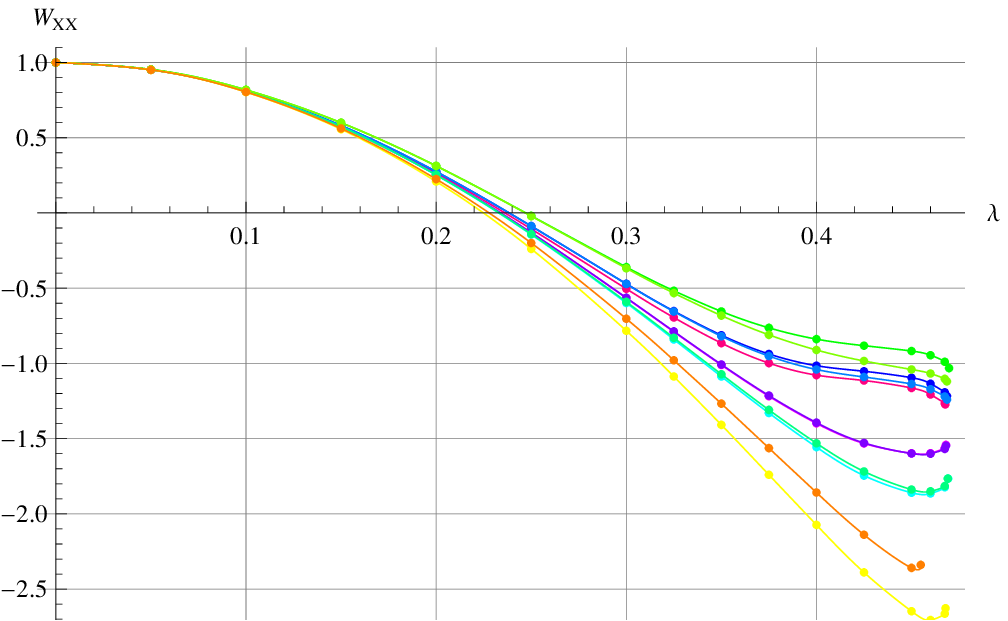}
 \caption{$W_{XX}$ for the marginal branch at levels $L=2,\ldots,12$.}
 \end{figure}

\newpage

{\footnotesize{
\begin{equation}
\begin{array}{|l|llllllll|}\hline
 L/\lambda & 0 & 0.05 & 0.1 & 0.15 & 0.2 & 0.25 & 0.3 & 0.325 \\\hline
 4  & 1 & 0.953629 & 0.816342 & 0.593424 & 0.29318  & -0.073308 & -0.491522 & -0.713913 \\
 5  & 1 & 0.953623 & 0.81625  & 0.592937 & 0.291551 & -0.07754  & -0.50075  & -0.726754 \\
 6  & 1 & 0.954787 & 0.807347 & 0.577341 & 0.26935  & -0.100336 & -0.509693 & -0.720384 \\
 7  & 1 & 0.95479  & 0.807286 & 0.57709  & 0.268623 & -0.101967 & -0.512727 & -0.72423  \\
 8  & 1 & 0.951803 & 0.808898 & 0.576478 & 0.263759 & -0.115659 & -0.519682 & -0.731568 \\
 9  & 1 & 0.951802 & 0.808879 & 0.576382 & 0.263459 & -0.116387 & -0.521665 & -0.733233 \\
 10 & 1 & 0.95142  & 0.808035 & 0.576461 & 0.261433 & -0.114106 & -0.526226 & -0.736616 \\
 11 & 1 & 0.951419 & 0.808023 & 0.576385 & 0.26121  & -0.114446 & -0.527015 & -0.738181 \\
 12 & 1 & 0.951444 & 0.807564 & 0.574423 & 0.264885 & -0.117854 & -0.529692 & -0.734165 \\\hline
\end{array}\nonumber
\end{equation}}}
{\footnotesize{
\begin{equation}
\begin{array}{|l|lllllll|}\hline
L/\lambda & 0.35 & 0.375 & 0.4 & 0.425 & 0.45 & 0.46 & 0.4675 \\\hline
4  & -0.940565 & -1.16652 & -1.38471 & -1.58423 & -1.74414 & -1.78381 & -1.78005 \\
5  & -0.95765  & -1.18798 & -1.40947 & -1.60904 & -1.76267 & -1.79768 & -1.79456 \\
6  & -0.928999 & -1.12958 & -1.31443 & -1.47336 & -1.59101 & -1.61814 & -1.61616 \\
7  & -0.93354  & -1.1343  & -1.31804 & -1.47347 & -1.58483 & -1.60904 & -1.60543 \\
8  & -0.933027 & -1.11522 & -1.26675 & -1.37868 & -1.4516  & -1.47037 & -1.47305 \\
9  & -0.934061 & -1.11437 & -1.26179 & -1.36752 & -1.43568 & -1.45435 & -1.45846 \\
10 & -0.931459 & -1.08781 & -1.21579 & -1.30324 & -1.35948 & -1.37767 & -1.38591 \\
11 & -0.930263 & -1.085   & -1.20917 & -1.29144 & -1.34508 & -1.36402 & -1.37425 \\
12 & -0.923884 & -1.08608 & -1.20608 & -1.28016 & -1.33251 & -1.35569 & -1.3723 \\\hline
\end{array}\nonumber
\end{equation}}}

\begin{figure}[ht]
 \centering
 \includegraphics[width=11cm]{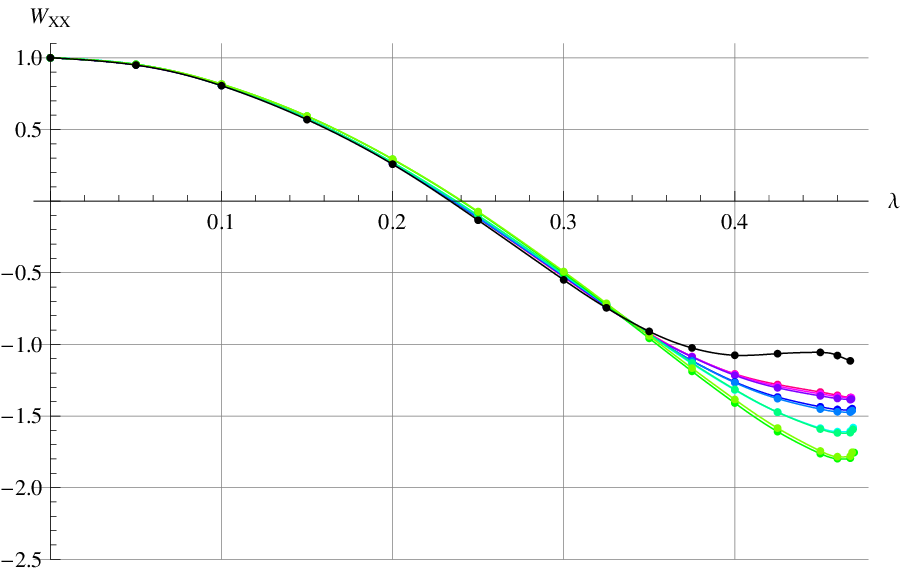}
 \caption{Pad\'{e}-Borel improvement of $W_{XX}$ for the marginal branch at levels $L=4,\ldots,12$.
 The black curve shows the pointwise extrapolation to the infinite level.}
 \end{figure}

{\footnotesize{
\begin{equation}
\begin{array}{|l|llllllll|}\hline
 L/\lambda& 0 & 0.05 & 0.1 & 0.15 & 0.2 & 0.25 & 0.3 & 0.325 \\\hline
 1  & 0.315384  & 0.329979  & 0.37345   & 0.444823 & 0.542391 & 0.663537 & 0.804395 & 0.880485 \\
 2  & 0.0406234 & 0.0572608 & 0.106764  & 0.18788  & 0.298435 & 0.435177 & 0.593475 & 0.678748 \\
 3  & 0.0406234 & 0.0561167 & 0.102266  & 0.178062 & 0.28175  & 0.410703 & 0.561182 & 0.642914 \\
 4  & 0.0121782 & 0.0277856 & 0.0742714 & 0.150608 & 0.255009 & 0.384804 & 0.536189 & 0.61837 \\
 5  & 0.0121782 & 0.0276126 & 0.0735889 & 0.149108 & 0.252436 & 0.38098  & 0.531054 & 0.612606 \\
 6  & 0.0048229 & 0.0202735 & 0.0662973 & 0.141892 & 0.245318 & 0.373974 & 0.524159 & 0.60576 \\
 7  & 0.0048229 & 0.020213  & 0.0660582 & 0.141365 & 0.24441  & 0.372619 & 0.52233  & 0.603702 \\
 8  & 0.0020698 & 0.0174607 & 0.0633076 & 0.138615 & 0.241657 & 0.369853 & 0.519535 & 0.600883 \\
 9  & 0.0020698 & 0.0174318 & 0.0631933 & 0.138363 & 0.24122  & 0.369198 & 0.518647 & 0.599882 \\
 10 & 0.0008175 & 0.0161771 & 0.0619308 & 0.137086 & 0.23992  & 0.367862 & 0.517257 & 0.598456 \\
 11 & 0.0008175 & 0.0161608 & 0.061866  & 0.136942 & 0.239671 & 0.367487 & 0.516748 & 0.597881 \\
 12 & 0.0001777 & 0.0155179 & 0.0612139 & 0.136273 & 0.238977 & 0.366755 & 0.515962 & 0.597061 \\\hline
\end{array}\nonumber
\end{equation}}}
{\footnotesize{
\begin{equation}
\begin{array}{|l|lllllll|}\hline
 L/\lambda& 0.35 & 0.375 & 0.4 & 0.425 & 0.45 & 0.46 & 0.4675 \\\hline
 1  & 0.95902  & 1.03852  & 1.11653  & -        & -       & -       & - \\
 2  & 0.766716 & 0.856006 & 0.944791 & 1.03037  & 1.10745 & -       & - \\
 3  & 0.72786  & 0.814928 & 0.902695 & 0.989157 & 1.071   & 1.10095 & 1.12114 \\
 4  & 0.703742 & 0.791192 & 0.879266 & 0.965916 & 1.04775 & 1.07762 & 1.0978 \\
 5  & 0.697405 & 0.784373 & 0.872105 & 0.958624 & 1.04067 & 1.07082 & 1.09143 \\
 6  & 0.690596 & 0.777582 & 0.865298 & 0.951738 & 1.03357 & 1.06352 & 1.08384 \\
 7  & 0.688329 & 0.775138 & 0.86273  & 0.949121 & 1.03102 & 1.06106 & 1.08148 \\
 8  & 0.685475 & 0.772236 & 0.859756 & 0.946035 & 1.02772 & 1.0576  & 1.07778 \\
 9  & 0.684371 & 0.771044 & 0.858503 & 0.944762 & 1.02649 & 1.05641 & 1.07662 \\
 10 & 0.6829   & 0.769518 & 0.856904 & 0.94306  & 1.02462 & 1.05441 & 1.07443 \\
 11 & 0.682265 & 0.768832 & 0.856183 & 0.942329 & 1.02391 & 1.05373 & 1.07376 \\
 12 & 0.681404 & 0.767921 & 0.855208 & 0.941268 & 1.02271 & 1.05243 & 1.07229 \\\hline
\end{array}\nonumber
\end{equation}}}

\begin{figure}[ht]
 \centering
 \includegraphics[width=11cm]{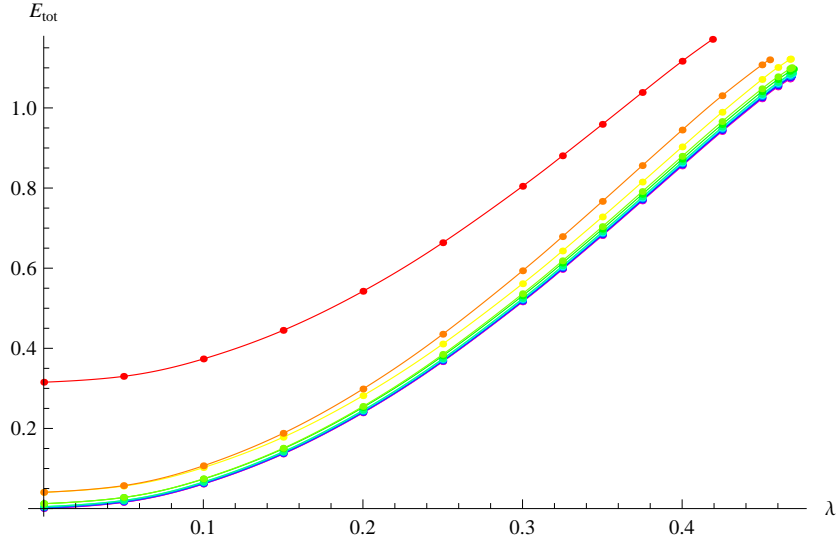}
 \caption{$E_{tot}$ for the vacuum branch at levels $L=1,\ldots,12$.}
 \end{figure}

\newpage

{\footnotesize{
\begin{equation}
\begin{array}{|l|llllllll|}\hline
 L/\lambda& 0 & 0.05 & 0.1 & 0.15 & 0.2 & 0.25 & 0.3 & 0.325 \\\hline
 1  & 0.315384  & 0.320266   & 0.335022  & 0.36      & 0.395882  & 0.443905 & 0.506437 & 0.544583 \\
 2  & 0.0406234 & 0.0461918  & 0.0630362 & 0.0915988 & 0.132712  & 0.187795 & 0.259302 & 0.302519 \\
 3  & 0.0406234 & 0.0458061  & 0.0614663 & 0.087961  & 0.125963  & 0.17662  & 0.241908 & 0.281081 \\
 4  & 0.0121782 & 0.0173993  & 0.0331765 & 0.0598732 & 0.0981732 & 0.149243 & 0.215096 & 0.254627 \\
 5  & 0.0121782 & 0.0173411  & 0.0329406 & 0.0593293 & 0.0971705 & 0.147593 & 0.212543 & 0.25149 \\
 6  & 0.0048229 & 0.00999118 & 0.0256074 & 0.0520254 & 0.0899106 & 0.140397 & 0.20544  & 0.244454 \\
 7  & 0.0048229 & 0.00997086 & 0.0255251 & 0.0518362 & 0.0895621 & 0.139823 & 0.204548 & 0.243353 \\
 8  & 0.0020698 & 0.00721809 & 0.0227735 & 0.049087  & 0.0868182 & 0.137089 & 0.201835 & 0.24066 \\
 9  & 0.0020698 & 0.0072084  & 0.0227343 & 0.0489971 & 0.0866528 & 0.136817 & 0.201409 & 0.240132 \\
 10 & 0.0008175 & 0.00595534 & 0.021479  & 0.0477386 & 0.085391  & 0.135553 & 0.200149 & 0.238879 \\
 11 & 0.0008175 & 0.00594987 & 0.0214569 & 0.047688  & 0.0852981 & 0.1354   & 0.199909 & 0.23858 \\
 12 & 0.0001777 & 0.00530908 & 0.0208133 & 0.0470401 & 0.084645  & 0.134742 & 0.199249 & 0.237923 \\\hline
\end{array}\nonumber
\end{equation}}}
{\footnotesize{
\begin{equation}
\begin{array}{|l|lllllll|}\hline
L/\lambda & 0.35 & 0.375 & 0.4 & 0.425 & 0.45 & 0.46 & 0.4675 \\\hline
1  & 0.588887 & 0.642031 & 0.711679 & -        & -        & -        & - \\
2  & 0.351907 & 0.40904  & 0.476817 & 0.561987 & 0.69613  & -        & - \\
3  & 0.325539 & 0.376472 & 0.435932 & 0.50811  & 0.605178 & 0.662188 & 0.743998 \\
4  & 0.299516 & 0.350982 & 0.411133 & 0.48427  & 0.582767 & 0.640283 & 0.71504 \\
5  & 0.295674 & 0.346272 & 0.405313 & 0.476921 & 0.572739 & 0.627705 & 0.692817 \\
6  & 0.288728 & 0.339459 & 0.398721 & 0.470764 & 0.567818 & 0.624362 & 0.695134 \\
7  & 0.287373 & 0.337786 & 0.396641 & 0.468137 & 0.564356 & 0.620319 & 0.689695 \\
8  & 0.28471  & 0.335177 & 0.394136 & 0.465871 & 0.562898 & 0.620073 & 0.695342 \\
9  & 0.284057 & 0.334364 & 0.393115 & 0.464569 & 0.561196 & 0.618168 & 0.693439 \\
10 & 0.282817 & 0.333153 & 0.391966 & 0.463577 & 0.560778 & 0.618688 & 0.700398 \\
11 & 0.282444 & 0.332686 & 0.391375 & 0.462815 & 0.559781 & 0.617605 & 0.699945 \\
12 & 0.281795 & 0.332055 & 0.390786 & 0.462338 & 0.559735 & 0.618308 & 0.708963 \\\hline
\end{array}\nonumber
\end{equation}}}

\begin{figure}[ht]
 \centering
 \includegraphics[width=11cm]{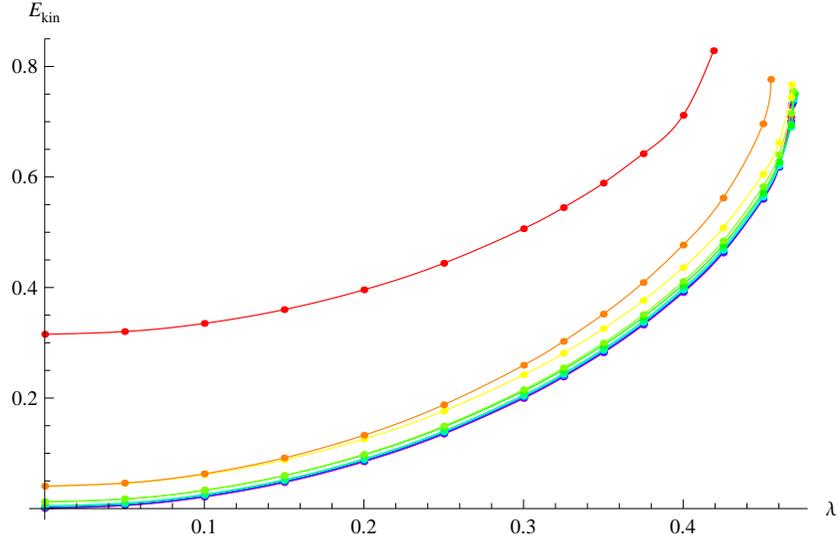}
 \caption{$E_{kin}$ for the vacuum branch at levels $L=1,\ldots,12$.}
 \end{figure}

\newpage
{\footnotesize{
\begin{equation}
\begin{array}{|l|llllllll|}\hline
 L/\lambda & 0 & 0.05 & 0.1 & 0.15 & 0.2 & 0.25 & 0.3 & 0.325 \\\hline
1  & 0.283437  & 0.285997  & 0.293789  & 0.30718   & 0.326882  & 0.35419  & 0.391582 & 0.415566 \\
2  & 0.110138  & 0.113727  & 0.124619  & 0.143208  & 0.170241  & 0.207002 & 0.25574  & 0.285807 \\
3  & 0.110138  & 0.113299  & 0.122891  & 0.139264  & 0.163072  & 0.195438 & 0.238292 & 0.264667 \\
4  & 0.0680476 & 0.0714674 & 0.0818461 & 0.0995642 & 0.125339  & 0.160392 & 0.206829 & 0.235422 \\
5  & 0.0680476 & 0.0713873 & 0.0815236 & 0.0988297 & 0.124009  & 0.158261 & 0.203649 & 0.231603 \\
6  & 0.0489211 & 0.0523251 & 0.0626591 & 0.0803113 & 0.106014  & 0.141016 & 0.187472 & 0.216127 \\
7  & 0.0489211 & 0.0522914 & 0.0625236 & 0.0800035 & 0.105459  & 0.14013  & 0.186158 & 0.214556 \\
8  & 0.0388252 & 0.0422404 & 0.05261   & 0.0703287 & 0.0961407 & 0.131317 & 0.178048 & 0.206902 \\
9  & 0.0388252 & 0.0422213 & 0.0525333 & 0.0701549 & 0.0958282 & 0.13082  & 0.177316 & 0.206028 \\
10 & 0.0318852 & 0.0353054 & 0.0456915 & 0.0634425 & 0.0893108 & 0.124581 & 0.171472 & 0.200444 \\
11 & 0.0318852 & 0.0352931 & 0.0456419 & 0.0633303 & 0.0891098 & 0.124263 & 0.171004 & 0.199886 \\
12 & 0.0274405 & 0.0308689 & 0.0412807 & 0.0590784 & 0.0850208 & 0.120403 & 0.167463 & 0.196553 \\\hline
\end{array}\nonumber
\end{equation}}}
{\footnotesize{
\begin{equation}
\begin{array}{|l|lllllll|}\hline
 L/\lambda &0.35 & 0.375 & 0.4 & 0.425 & 0.45 & 0.46 & 0.4675 \\\hline
 1  &  0.444721 & 0.481853 & 0.534983 & -        & -        & -        & - \\
 2  &  0.320791 & 0.362181 & 0.412733 & 0.47892  & 0.591238 & -        & - \\
 3  &  0.295248 & 0.33119  & 0.374481 & 0.429179 & 0.507074 & 0.555604 & 0.629843 \\
 4  &  0.268586 & 0.307574 & 0.35454  & 0.413848 & 0.497985 & 0.549673 & 0.62015 \\
 5  &  0.264032 & 0.30216  & 0.348087 & 0.406038 & 0.487846 & 0.537216 & 0.598286 \\
 6  &  0.249417 & 0.288628 & 0.335981 & 0.395973 & 0.481405 & 0.533842 & 0.602502 \\
 7  &  0.247551 & 0.286425 & 0.333385 & 0.392901 & 0.477692 & 0.529701 & 0.597151 \\
 8  &  0.240452 & 0.280018 & 0.327884 & 0.388705 & 0.475913 & 0.530177 & 0.605092 \\
 9  &  0.239417 & 0.278801 & 0.326456 & 0.387033 & 0.473965 & 0.52814  & 0.603253 \\
 10 &  0.234153 & 0.273942 & 0.322141 & 0.38353  & 0.472071 & 0.527904 & 0.610834 \\
 11 &  0.233494 & 0.273169 & 0.321237 & 0.382477 & 0.470869 & 0.526706 & 0.610477 \\
 12 &  0.230412 & 0.270402 & 0.318887 & 0.380741 & 0.470354 & 0.527499 & 0.621024 \\\hline
\end{array}\nonumber
\end{equation}}}

\begin{figure}[hb]
 \centering
 \includegraphics[width=11cm]{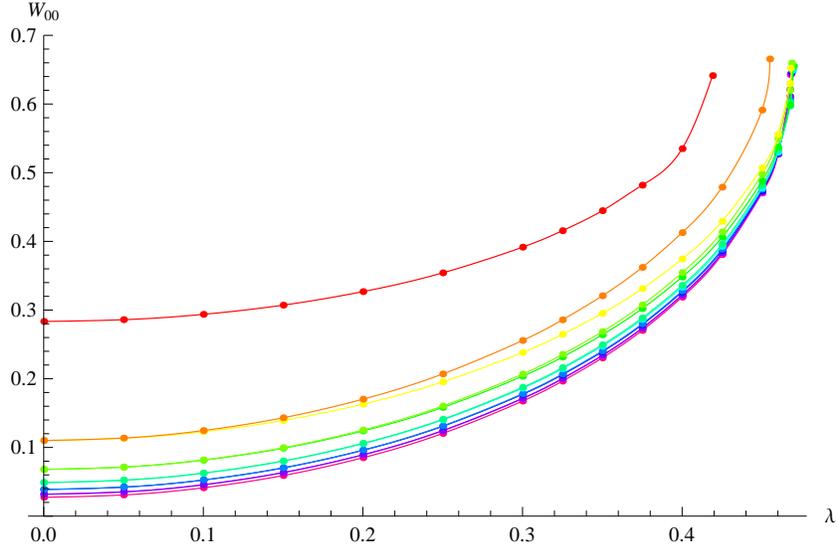}
 \caption{$W_{00}$ for the vacuum branch at levels $L=1,\ldots,12$.}
 \end{figure}

\newpage

{\footnotesize{
\begin{equation}
\begin{array}{|l|llllllll|}\hline
 L/\lambda& 0 & 0.05 & 0.1 & 0.15 & 0.2 & 0.25 & 0.3 & 0.325 \\\hline
 1  & 0.283437  & 0.285997  & 0.293789   & 0.30718    & 0.326882  & 0.35419   & 0.391582  & 0.415566 \\
 2  & 0.110138  & 0.0875615 & 0.0193784  & -0.0958322 & -0.260671 & -0.479375 & -0.758784 & -0.92461 \\
 3  & 0.110138  & 0.0893906 & 0.0265967  & -0.0799804 & -0.233546 & -0.439388 & -0.706161 & -0.866689 \\
 4  & 0.0680476 & 0.0496969 & -0.0050390 & -0.0951681 & -0.218875 & -0.373213 & -0.553475 & -0.651065 \\
 5  & 0.0680476 & 0.0501372 & -0.0033118 & -0.0914059 & -0.21248  & -0.363751 & -0.540592 & -0.636278 \\
 6  & 0.0489211 & 0.0313177 & -0.021478  & -0.109444  & -0.232628 & -0.391325 & -0.586511 & -0.69853 \\
 7  & 0.0489211 & 0.0315214 & -0.0206724 & -0.107665  & -0.229551 & -0.386693 & -0.580163 & -0.691308 \\
 8  & 0.0388252 & 0.0215946 & -0.0299634 & -0.115428  & -0.234027 & -0.384488 & -0.564713 & -0.664932 \\
 9  & 0.0388252 & 0.0217196 & -0.0294676 & -0.114329  & -0.23211  & -0.381561 & -0.560599 & -0.66015 \\
 10 & 0.0318852 & 0.0149249 & -0.0358963 & -0.120404  & -0.238325 & -0.389336 & -0.573202 & -0.677496 \\
 11 & 0.0318852 & 0.0150087 & -0.0355634 & -0.119663  & -0.23703  & -0.387355 & -0.570422 & -0.674281 \\
 12 & 0.0274405 & 0.01062   & -0.0397397 & -0.12332   & -0.239541 & -0.387463 & -0.565568 & -0.665187 \\\hline
\end{array}\nonumber
\end{equation}}}
{\footnotesize{
\begin{equation}
\begin{array}{|l|lllllll|}\hline
 L/\lambda & 0.35 & 0.375 & 0.4 & 0.425 & 0.45 & 0.46 & 0.4675 \\\hline
 1  & 0.444721  & 0.481853  & 0.534983  & -        & -        & -        & - \\
 2  & -1.11066  & -1.32032  & -1.55916  & -1.83858 & -2.20227 & -        & - \\
 3  & -1.04892  & -1.25712  & -1.49816  & -1.78477 & -2.14961 & -2.3441  & -2.57644 \\
 4  & -0.751803 & -0.853628 & -0.953435 & -1.04596 & -1.11977 & -1.13655 & -1.13214 \\
 5  & -0.734843 & -0.83396  & -0.929944 & -1.01617 & -1.07704 & -1.08354 & -1.06612 \\
 6  & -0.82101  & -0.955211 & -1.10355  & -1.27139 & -1.47483 & -1.58038 & -1.69678 \\
 7  & -0.812938 & -0.946358 & -1.09407  & -1.26161 & -1.46574 & -1.57244 & -1.69065 \\
 8  & -0.770984 & -0.881767 & -0.995469 & -1.10865 & -1.21245 & -1.24455 & -1.25248 \\
 9  & -0.765464 & -0.875398 & -0.988039 & -1.09967 & -1.20044 & -1.22985 & -1.23252 \\
 10 & -0.790205 & -0.911714 & -1.04293  & -1.18621 & -1.34996 & -1.42987 & -1.51928 \\
 11 & -0.786535 & -0.90757  & -1.03829  & -1.18108 & -1.34438 & -1.42422 & -1.51448 \\
 12 & -0.771212 & -0.882866 & -0.998863 & -1.1167  & -1.22961 & -1.26769 & -1.27831 \\\hline
\end{array}\nonumber
\end{equation}}}

\begin{figure}[hb]
 \centering
 \includegraphics[width=11cm]{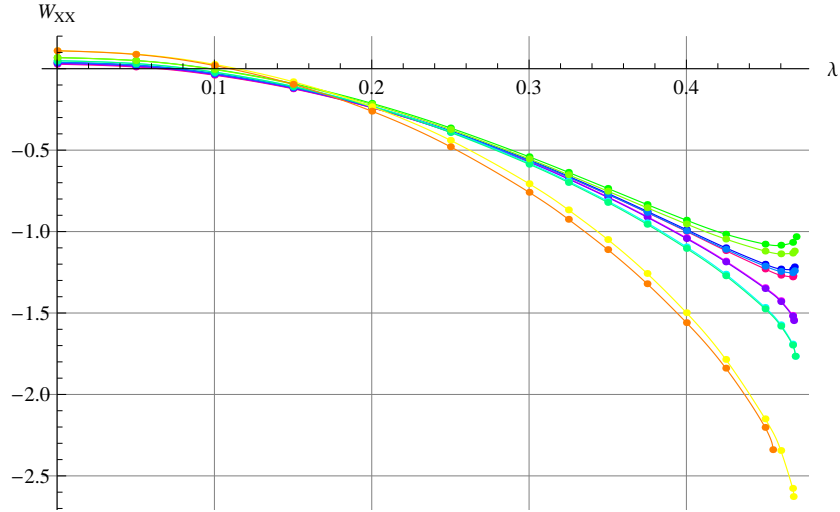}
 \caption{$W_{XX}$ for the vacuum branch at levels $L=2,\ldots,12$.}
 \end{figure}

\newpage
{\footnotesize{
\begin{equation}
\begin{array}{|l|llllllll|}\hline
\! L/\lambda \! & 0 & 0.05 & 0.1 & 0.15 & 0.2 & 0.25 & 0.3 & 0.325 \\\hline
 4  & 0.0531112 & 0.0440065  & -0.00531479 & -0.0958717 & -0.223919 & -0.388704 & -0.590782 & -0.706525 \\
 5  & 0.0531112 & 0.0445574  & -0.0035379 & -0.0922339 & -0.217909 & -0.379966 & -0.579115 & -0.69336 \\
 6  & 0.0390372 & 0.0220318  & -0.0257739 & -0.109098 & -0.230925 & -0.388482 & -0.580837 & -0.689993 \\
 7  & 0.0390372 & 0.0222627  & -0.0248376 & -0.10725 & -0.227899 & -0.384041 & -0.574815 & -0.683133 \\
 8  & 0.0309852 & 0.0160397  & -0.034642 & -0.118676 & -0.234626 & -0.392628 & -0.574418 & -0.672585 \\
 9  & 0.0309852 & 0.0161757  & -0.034102 & -0.117464 & -0.232628 & -0.390156 & -0.570038 & -0.667684 \\
 10 & 0.0286352 & 0.0120373  & -0.0381778 & -0.120848 & -0.234825 & -0.388183 & -0.570068 & -0.671444 \\
 11 & 0.0286352 & 0.0121251  & -0.0378259 & -0.120048 & -0.233877 & -0.386267 & -0.567252 & -0.668102 \\
 12 & 0.0269078 & 0.00947608 & -0.0405434 & -0.123708 & -0.239038 & -0.388482 & -0.566975 & -0.66749 \\\hline
\end{array}\nonumber
\end{equation}}}
{\footnotesize{
\begin{equation}
\begin{array}{|l|lllllll|}\hline
L/\lambda & 0.35 & 0.375 & 0.4 & 0.425 & 0.45 & 0.46 & 0.4675 \\\hline
4  & -0.83294  & -0.971294 & -1.12392 & -1.29564 & -1.50001 & -1.60228 & -1.71059 \\
5  & -0.818275 & -0.955145 & -1.10633 & -1.27674 & -1.48004 & -1.5817  & -1.68443 \\
6  & -0.80794  & -0.935052 & -1.07225 & -1.22189 & -1.39179 & -1.47265 & -1.5529 \\
7  & -0.800211 & -0.926413 & -1.06263 & -1.21114 & -1.37956 & -1.4596  & -1.53858 \\
8  & -0.776614 & -0.889276 & -1.01512 & -1.1567  & -1.3118  & -1.37959 & -1.44152 \\
9  & -0.771045 & -0.882701 & -1.00728 & -1.14783 & -1.302   & -1.36894 & -1.42939 \\
10 & -0.777659 & -0.889965 & -1.01008 & -1.13751 & -1.27054 & -1.32578 & -1.37297 \\
11 & -0.773676 & -0.885449 & -1.00499 & -1.13172 & -1.26362 & -1.31793 & -1.36347 \\
12 & -0.775368 & -0.895805 & -1.00836 & -1.136   & -1.26819 & -1.32277 & -1.36773 \\\hline
\end{array}\nonumber
\end{equation}}}

\begin{figure}[ht]
 \centering
 \includegraphics[width=11cm]{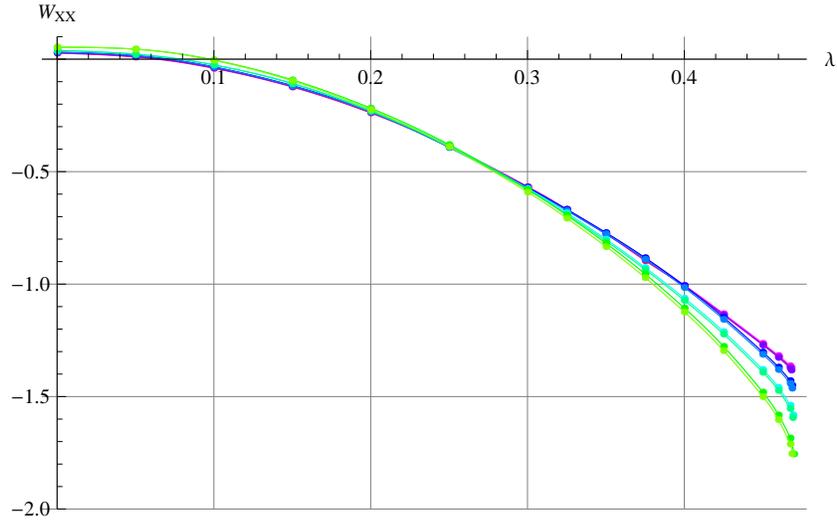}
 \caption{Pad\'{e}-Borel improvement of $W_{XX}$ for the vacuum branch at levels $L=4,\ldots,12$.}
 \end{figure}

\begin{figure}[ht]
 \centering
 \includegraphics[width=5cm]{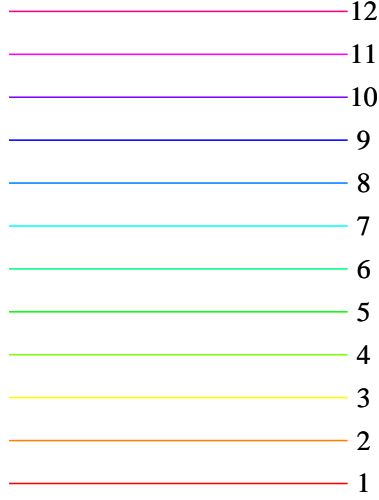}
 \caption{Colors of different levels in plots.}
\label{Fig:legend}
 \end{figure}

\normalsize
\begin{table}[!t]
\bigskip\bigskip
\begin{equation}
\begin{array}{|l|llll|}\hline
 L & E_{tot} & E_{kin} & W_{00} & W_{XX} \\\hline
 1  & 1.17115 & 0.828479 & 0.641335 & 0.641335 \\
 2  & 1.12004 & 0.77663  & 0.665624 & -2.33899 \\
 3  & 1.12209 & 0.767249 & 0.652168 & -2.62701 \\
 4  & 1.10045 & 0.754992 & 0.659607 & -1.11941 \\
 5  & 1.09696 & 0.750709 & 0.655268 & -1.03139 \\
 6  & 1.08768 & 0.744818 & 0.652912 & -1.7654  \\
 7  & 1.08567 & 0.740952 & 0.649266 & -1.76564 \\
 8  & 1.0805  & 0.737844 & 0.649236 & -1.24035 \\
 9  & 1.07929 & 0.735213 & 0.646753 & -1.21667 \\
 10 & 1.07597 & 0.732495 & 0.644782 & -1.54666 \\
 11 & 1.07518 & 0.73124  & 0.643665 & -1.54135 \\
 12 & 1.07286 & 0.729149 & 0.642705 & -1.27262 \\\hline
\end{array}\nonumber
\end{equation}
\caption{The energy and the gauge-invariant observables
of solutions constructed exactly at $\lambda_{crit}$ for each given level.}
\label{t-critical}
\end{table}

\end{document}